\documentclass[onecolumn,english]{aa}
\usepackage{amsmath}
\PassOptionsToPackage{version=3}{mhchem}
\usepackage{mhchem}
\usepackage{esint}
\usepackage{graphicx}
\usepackage{mwe}
\usepackage{booktabs} % For prettier tables
\newcommand{\teff}{T$_{\!e\!f\!f}$} 
\newcommand{\tenv}{T$_{env}$}  
\newcommand{\renv}{R$_{env}$}  
\usepackage[pdftex]{adjustbox}

\newcommand{\shortunderscore }{\kern-.5pt\adjustbox{scale={0.5}{1},raise={1.5pt}{\height}}{\textunderscore}\kern-.5pt}

\usepackage{babel}
\bibpunct[, ]{(}{)}{;}{a}{,}{,}

\begin{document}

\title{Spectral line fluorescence in moving envelopes of stars}
\subtitle{I. Theoretical background, numerical code, and proof of concept}
\author{Claude Bertout\thanks{Retired. Current e-mail: claude.m.bertout (at) gmail.com}}
\institute{\center{Institut d'Astrophysique de Paris\\98bis, Boulevard Arago, 75014 Paris, France}}
\date{Received February 29, 2024; accepted May 28, 2024}

\abstract
%context
{Fluorescence of the optical \ce{FeI} lines is observed in active T~Tauri stars, and is considered a defining characteristic of this class of young stellar objects. }
%aims
{The formation of optical fluorescent lines in moving media has not yet been studied in detail. This work represents a first step in investigating the fluorescence process in different types of macroscopic velocity fields: (a) accelerated outflows, (b) accelerated infalls, and (c) non-monotonic velocity fields (such as an accelerating outflow followed by a deceleration region or an accretion shock front). We aim to develop a general computer code for studying the fluorescent emission in any 2D macroscopic velocity field. As an illustration, we investigate \ce{FeI} T~Tauri-like fluorescent emission in these moving stellar envelopes.}
%methods
{We solved the radiative transfer equations for the lines involved in the fluorescent process, assuming spherical symmetry and a simplified atomic model. We used the framework of the generalized Sobolev theory to compute the interacting, nonlocal source functions. The emergent line fluxes were then integrated exactly. }
%results
{Because of Doppler shifts in the moving gaseous envelope, photons of the three lines involved in T~Tauri star \ce{FeI} fluorescence (\ce{CaII} H, \ce{FeI}$\lambda3969$, and \ce{H}$_\epsilon$) interact with each other in a complex way, such that fluorescent amplification of the line flux occurs not only for \ce{FeI}$\lambda3969$, but also for the other two lines, in all velocity fields that we investigated. 
With the assumption of local thermodynamic equilibrium  populations, the line source functions of moderately optically thick lines are not strongly affected by line interactions, while they are depressed in the inner envelope for optically thick lines because of stellar photon absorption in the interaction regions. Fluorescent amplification takes place mainly in the observer's reference frame during the flux integration. 
We define a measure of fluorescence based on the line equivalent widths and perform a parameter study for an accretion flow with a broad range of envelope temperatures and densities while including approximate collisional de-excitation rates in the source function computations. Significant fluorescence occurs over the entire temperature range of the  investigated flow, but only in the higher density range, suggesting that relatively high mass accretion rates are needed to trigger the fluorescence process.}
%conclusions
{Further comparison with observations will require solving the rate equations for the atomic populations involved, along with the radiation field computed with the method presented here. The main product of this research is the open-source computer code \textit {SLIM2} (Spectral Line Interactions in Moving Media), written in Python/NumPy, which numerically solves the fluorescence problem for arbitrary 2D velocities.}

        
\maketitle

\section{Introduction}\label{sec:introduction}

The physical process of fluorescence, also known as photoexcitation by accidental resonance, is associated in stellar astronomy with the name Ira S. Bowen, who showed that the spectra and composition of gaseous nebulae could be explained by fluorescent emission of the ions \ce{OIII} and \ce{NIII} excited by neighboring \ce{HeII} transitions \citep{1934PASP...46..146B, 1935ApJ....81....1B}. This solved a long-standing mystery that had even led \citet{1898ApJ.....8R..54H} to postulate the existence of a chemical element he  called ``nebuliu''. Fluorescent emission was also successful in explaining features of the line spectra in long-period variables \citep{1937ApJ....86..499T, 1972A&A....17..354W} as well as in symbiotic stars \citep[e.g.,][]{1991Ap&SS.185..265K,1999MNRAS.309.1074P, 2004RMxAC..21..132E, 2007A&A...464..715S, 2018JASS...35....7H}.

A similar mechanism was proposed by \citet{1945PASP...57..166H} to account for the anomalous \ce{Fe I} emission observed in active T~Tauri stars (TTSs). Specifically, the transitions $\lambda$4063 and $\lambda$4132 of \ce{Fe I} appear in strong emission in active TTSs, while other lines of the multiplet $a^3F - y^3F^0$ are in very weak emission or are not detected \citep{1945ApJ...102..168J}. The two anomalous lines originate from the same $y^3F^0_3$ upper level together with the $\lambda$3969 line. Herbig conjectured that ``the explanation lies in the near coincidence of the $\lambda$3969 line of \ce{Fe I} with the red component of the H line of \ce{Ca II}. Absorption of the proper frequencies by the \ce{Fe I} atoms would ensure a supply of atoms to $y^3F^0_3$, while the other sub-levels of $J=2\ \rm{and}\ 4$ would lack such an active source.'' The $\lambda$3969.3 line is flanked by \ce{Ca II} H at $\lambda$3968.5 on the blue side and by the \ce{H I} $H_\epsilon$ line at $\lambda$3970.1 on the red side. Both of these lines are in strong emission in TTSs, and given the complexity of motions in the surroundings of these stars, it is conceivable that either one could  be responsible for pumping atoms into the upper level of \ce{Fe I} $\lambda$3969. Herbig did not consider the possible role of H$_\epsilon$ in the \ce{Fe I} fluorescence, presumably because it was believed in 1945 that TTS activity was a scaled-up version of solar-type, chromospheric activity accompanied by a strong stellar wind.\ However, \citet{1974ApJ...191..143W, 1975ApJ...197..365W} concluded from a detailed investigation of the statistical equilibrium of the lines involved that both H and H$_\epsilon$ were contributing to the fluorescence mechanism in the active TTS RW~Aur, and he suggested that rotational velocities were greater than ejection velocities in the envelope of that star. While we know now that TTSs are not fast rotators, and that TTS activity is fueled mainly by accretion onto the star of disk material, it is certainly plausible that either H or H$_\epsilon$, or both, are at work in the fluorescence process. And given the wealth of mid- and high-resolution spectroscopic data that has become available with the advent of such instruments as ESO's ESPRESSO, UVES, and X-Shooter \citep [see, e.g.,][] {2023A&A...679A..14A}, it is obvious that a renewed quantitative study of fluorescence in young stars is overdue.
%\end{document}
Our aim is to present the theoretical background as well as a numerical code that enables such a study. The problem is relatively complex, and we will therefore solve it in two steps. 

The first one, presented in the present paper, focuses on the formation of the $\lambda$3969 \ce{FeI} line and its interactions with the \ce{CaII} H and \ce{HI} H$_\epsilon$ lines, using the equivalent two-level atom approach for each of the lines and the assumption of local thermodynamic equilibrium (LTE) for computing the populations of the atomic levels involved. Our aim is to create a code for solving the radiative interactions between lines that occur in the moving media that we consider and to gain some understanding of the physical conditions of the envelope that lead to amplification of the $\lambda$3969 line flux. 

This first investigation will not make a direct comparison between models and observations possible, since we will not be solving the statistical equilibrium equations for the level populations of \ce{FeI} self-consistently with the radiation field. In a forthcoming extension of this work, we will therefore couple the code developed here with a computation of the non-LTE rate equations for a multilevel \ce{Fe} atom model, which will then allow us to compare the anomalous $ \ce{FeI} \lambda$4063 and $\lambda$4132 lines, as well as the nonfluorescent lines of the same multiplet, to observations.

\medskip

\noindent
The physical processes taking place in the most active T~Tauri envelopes involve several components: circumstellar disk, magnetospheric accretion columns between disk and star, bipolar disk wind, and stellar wind. Modeling the line formation process in this dynamic environment is challenging, but considerable advances have been achieved over the past 20 years \citep[e.g.,][]{2001ARep...45..442T, 2014A&A...562A.104T, 2012MNRAS.426.2901K}. All of this work has been done in the framework of the generalized Sobolev approximation, which makes this problem tractable.

In a moment of remarkable insight, Sobolev (1947, English translation in \citeyear{1960mes..book.....S}) realized that the Doppler shift simplifies the formation of spectral lines in moving media such as stellar envelopes and winds. The basic approximation made in what is now called the Sobolev theory is that, because of the macroscopic velocity gradient in the moving gas, the source function and optical depth of a given transition are constant over the narrow region where emitted photons are not Doppler-shifted with respect to atoms in the rest of the envelope, and equal to zero everywhere else. 
This is true in particular in outwardly accelerating, supersonic winds, which represent the simplest case for applying the Sobolev theory; the line formation problem then reduces to finding regions of constant radial velocity with respect to the observer, each of which corresponds to a single frequency in the emergent line profile. 

In more quantitative terms, Sobolev assumed in his original derivation that the local absorption and emission profiles were constant over the width $\Delta\nu_{ik}=2(u/c)\nu_{ik}$ and equal to zero everywhere else. Here, $\nu_{ik}$ is the frequency of transition $ik$, $u$ is the mean thermal velocity of atoms, and $c$ is the velocity of light. For a hydrogen line at $10^{4}$K, the emission and absorption profile widths are on the order of 10~km/s, so $\Delta\nu_{ik}/\nu_{ik}$ is on the order of $10^{-4}$, or 0.5Å\ for a line wavelength of 5000Å. This is to be compared to the overall line width, which will extend, for example, over 16Å\ in a 500~km/s outflow. In other words, whenever the Sobolev approximation holds, absorption and emission line profiles can be seen as delta-like functions, which is consistent with considering that the emitting regions at a given line frequency are radial velocity surfaces rather than extended volumes.

The Sobolev theory has been used extensively for modeling line formation in envelopes of stars and outflows in a variety of objects, ranging from early-type stars \citep[e.g.,][]{1970MNRAS.149..111C,1987ApJ...314..726L} to young stellar objects \citep[e.g.,][]{2019AstL...45..371D} and from interstellar clouds \citep{1974ApJ...189..441G} to bipolar outflows \citep[e.g.,][]{1986ApJ...307..313C, 1990ApJ...348..530C, 1992A&A...261..274C}. A review of the rich literature applying Sobolev's theory in various contexts is beyond the scope of this paper, but it is fair to say that this approach has led to huge progress in our understanding of line formation in many different astrophysical situations, as Sobolev himself had foreseen in his original treatise.

Another reason for this success -- besides the simplification in line formation that we have alluded to -- stems from the fact that the Sobolev method was shown, by extensive comparisons with much more computationally challenging numerical solutions of the exact line transfer equations, to be sufficiently accurate for a large class of problems, especially when the approximation is used only for computing the line source function and the emergent line flux is integrated exactly \citep[see, e.g.,][]{1980A&A....86..105B, 1981A&A....93..353H}.

Originally used for the study of accelerating outflows, the Sobolev method has been generalized and extended over the years to more involved velocity fields, including decelerated outflows, accelerated inflows, rotation, and various combinations thereof. The generalization of the original Sobolev theory for treating velocity laws that lead to complex radial velocity surfaces was worked out by \citet{1978ApJ...219..654R} in a seminal paper. The same theoretical framework has also been useful for studying the interactions between lines that occur in some astrophysical situations, for example close doublets \citep[cf.][]{1982ApJ...255..267O} and masers \citep[cf.][]{1996ApJ...467..292P}. The latter authors considered the case of rectangular local emission and absorption profiles, in which case the line source function computation of the locally interacting lines becomes analytically tractable, albeit tedious.

In static media, spectral lines interact when their absorption profiles overlap locally. If macroscopic velocity gradients are present in the line-emitting region, however, it can also happen that the line of interest is Doppler-shifted onto another one that is too far in wavelength for a local interaction to occur; following \citet{1996ApJ...467..292P}, we call this situation a nonlocal interaction. 

Quantitatively, two lines, A and B, with rest wavelengths $\lambda_{A}$ and $\lambda_{B}$ and absorption profile widths $\Delta\lambda_{A}$ and $\Delta\lambda_{B}$ interact locally when $(\lambda_{B}-\lambda_{A})/\lambda_{A} \lesssim \Delta\lambda_{A} + \Delta\lambda_{B}$. They interact non-locally whenever 
line-A photons emitted by matter moving at velocity $v_{A}$ cross a distant region where the radial velocity, $v_{r}^{B}$, of the B-line-emitting matter is such that $(\lambda_{B}-\lambda_{A})/\lambda_{A}  = (v_{r}^{B}-v_{r}^{A})/c$.\footnote{For the sake of simplicity, we neglected the line widths in this condition.} In the example discussed earlier, the line at 5000Å\ will interact to some extent with neighboring lines located within 8Å\ of the transition rest wavelength.

In the present work, we calculate the line source functions and emergent line profiles of radiatively overlapping lines in moving media. While the generalized Sobolev theory \`{a} la Hummer and Rybicki remains the conceptual framework underlying the computation, we make use of the Sobolev approximation for computing the source function and, following \citet{1984ApJ...285..269B}, perform an exact integration of the emerging flux, assuming Gaussian shapes for the local absorption profiles.  

\medskip

\noindent
We make several simplifying assumptions in the subsequent discussion. The moving medium under consideration is a spherically symmetric, extended stellar envelope with maximum radius \renv\  surrounding a star of mass $M_*$, radius $R_{c}$, and effective temperature \teff. The envelope velocity field is, at first, assumed to be a monotonically accelerating outflow from velocity $V_{c}$ at the stellar core to $V_{max}$ at \renv. We then consider velocity fields that lead to more complex distant radiative coupling. We neglect continuous opacity in the envelope, which has temperature T$_{env}(r)$. The gas number density, $n_{t}$, is computed from the continuity equation. The emitting gas has a solar composition, and the atomic level populations and free electron number density, $n_{e}(r)$, follow from the assumption of LTE.  In the framework of equivalent two-level atoms, $n_{l}$ and $n_{u}$, the occupation numbers of both line levels in Eq.~\ref{Eq.chi}, are then given by the Saha-Boltzmann equation evaluated at the local temperature and electron density. 

Collisional de-excitation of atoms is an important radiation source for the lines that we investigate, so we included approximate rates, $\epsilon_k$, in the source function computation for line $k$ . However, since most of the previous theoretical work done on line formation in moving media neglects the collisional term and since we want to be able to compare our results to previous work, we will assume $\epsilon = 0$  in the theoretical sections of this work and reserve the use of collisional terms to Sects.~\ref{sec:extension-to-complex-velocity-fields} and~\ref{sec:Discussion}.

In the next section, we study the fluorescence process in a monotonically  accelerating outflow. The Sobolev approximation for the source functions of interacting lines is discussed in Sect.~\ref{subsec:Sobolev-source-function}, while Sect.~\ref{sec:Exact-line-profile} extends Bertout's 1984 exact flux integration method to the case at hand. In Sect.~\ref{sec:extension-to-complex-velocity-fields} we generalize the analysis to the more complex radial velocity surfaces found in accreting envelopes (Sect.~\ref{subsec:accretion-flow}) and in non-monotonic velocity fields (Sect.~\ref{subsec:non-monotonic-flow}). A discussion of the physical conditions that lead to fluorescence follows in Sect~\ref{sec:Discussion}. The \ce{FeI} fluorescence observed in TTSs is used as an example in the various illustrations of the paper.

Finally, we briefly present in Appendix~\ref{sec:SLIM2} the open-source computer code (available on GitHub under CC 4.0 license) written in Python/NumPy that solves the nonlocal radiative coupling between interacting spectral lines formed in spherical, moving envelopes of stars.

\section{Formation of interacting lines in a monotonic, accelerated outflow\label{sec:accelerated-outflow}}

\begin{figure}
        \centering
        \begin{minipage}{0.48\textwidth}
                \centering
                \includegraphics[width=1.0\textwidth]{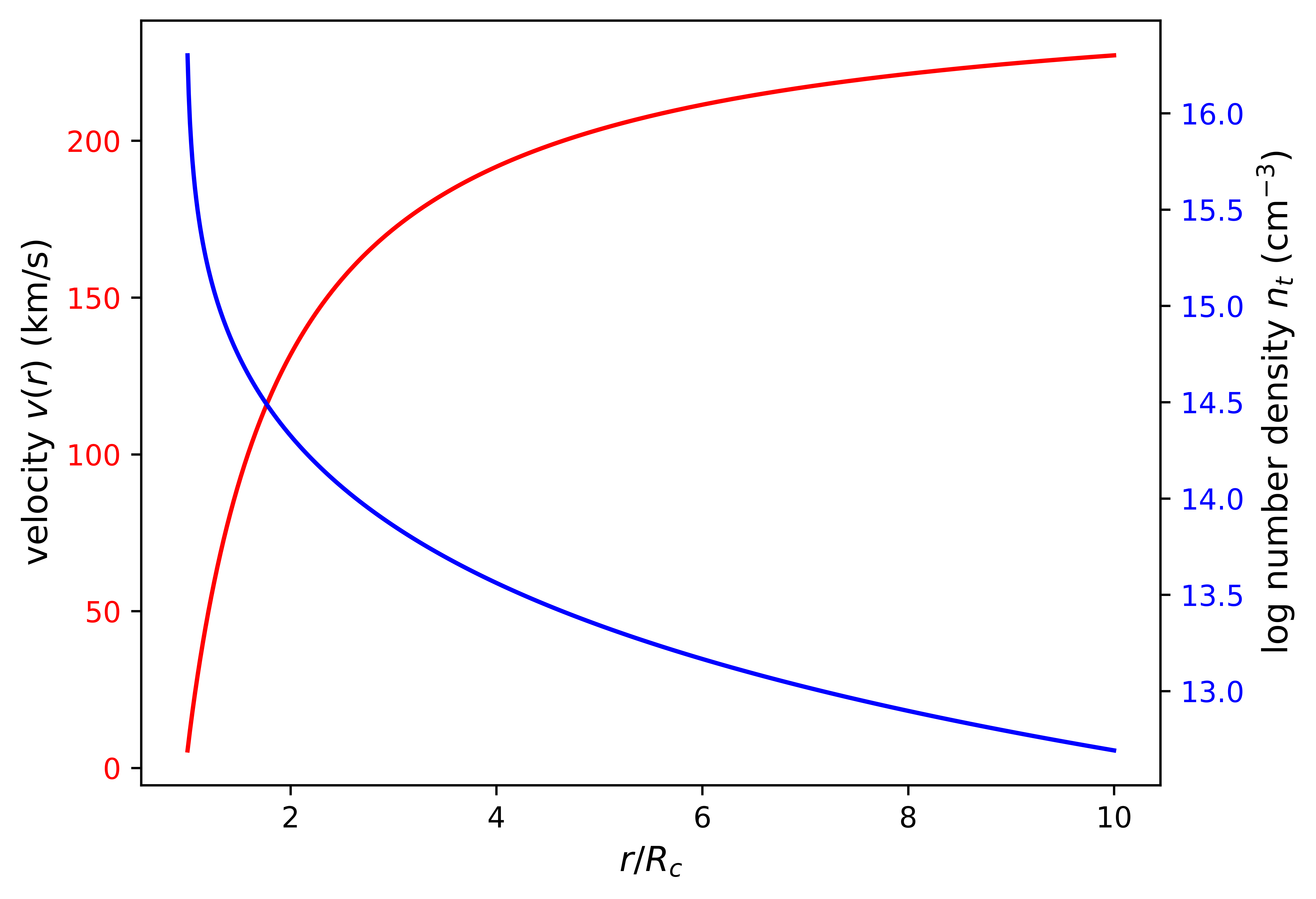} 
                \caption[Short caption]{Velocity (Eq.~\ref{eq:outflow}) and density law (Eq.~\ref{eq:density}) of the stellar outflow. }
                \label{fig:figure2.1}
        \end{minipage}\hfill
        \begin{minipage}{0.48\textwidth}
                \centering
                \includegraphics[width=0.9\textwidth]{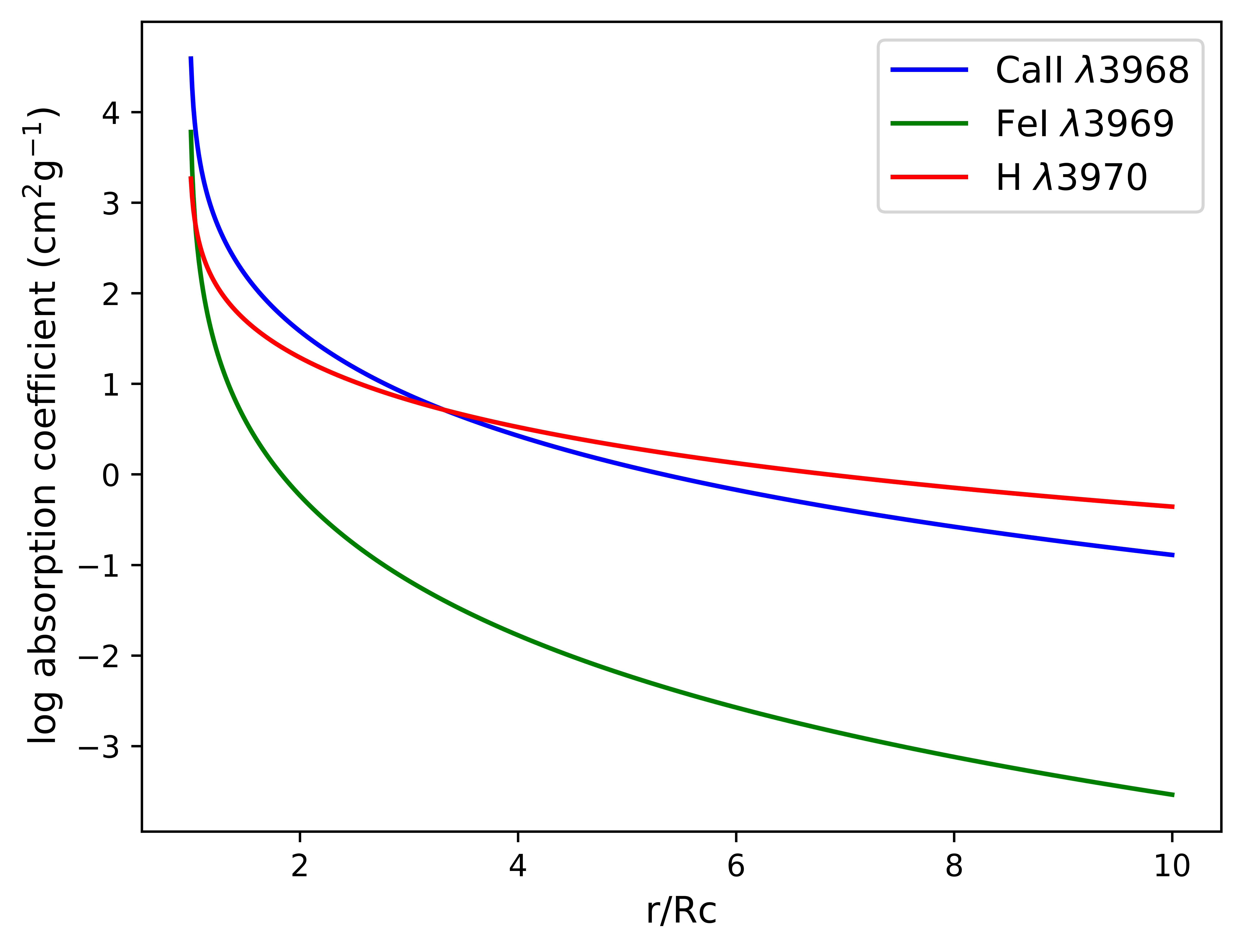} 
                \caption[Short caption]{Line absorption coefficient for the three lines considered.}
                \label{fig:figure2.2}
        \end{minipage}
\end{figure}

We considered an outwardly accelerating stellar outflow with properties loosely inspired by the Alfv\'en-wave-driven stellar wind originally proposed for TTSs by \citet{1982ApJ...261..279H}. At first, we used the data they give in their table for the velocity of Model 5, interpolating between the few provided radial points, but this resulted in discontinuous derivatives of the velocity at the grid points that in turn led to spurious variations in the source function of the optically thick hydrogen line. We therefore looked for a least-square fit to the data, assuming a parameterized velocity of the form
\begin{equation}
        v(r) = v_\infty\left [ 1 - (1-v_{min})\frac{r}{R_c}\right ]^{\alpha}.\label{eq:outflow}
\end{equation}
The resulting parameters are given in Table~\ref{tab:table1} together with the other computation parameters, and  Fig.~\ref{fig:figure2.1} shows the velocity law together with the adopted gas density law, which follows from the continuity equation
\begin{equation}
        n_t(r) = n_t(R_c) \frac{v(R_c)}{v(r)} \left(\frac{R_c}{r}\right)^2.\label{eq:density}
\end{equation} 
Our choice of computation parameters for the envelope properties is informed by spectroscopic data of the active TTS system SCrA  \citep [e.g., ][]{1982A&AS...47..419B}, which display the \ce{FeI} fluorescence phenomenon discussed here. In these spectra, the intensities of the \ce{CaII} H + \ce{H}$_\epsilon$ blend and the \ce{CaII} K line, both in strong emission, are of comparable strength. This is also the case, more generally, for many TTSs \citep{1993AJ....106.2024V}.  For this to occur (within the limitations of the LTE assumption), the temperature of the line formation region cannot be higher than $6 - 7 \cdot 10^3 $K. At such low temperatures, we must have a relatively high gas density in order to get significant emission line flux. In the range of values chosen here, the three lines are partially optically thick at the bottom of the envelope (Fig.~\ref{fig:figure2.2}). 
Clearly,  near-photospheric conditions are required for producing moderately strong emission lines when using the assumptions of LTE populations and isothermal gas, and when neglecting collisions. As a consequence, the mass-flow rate determined using these assumptions and the additional condition of spherical symmetry is much higher than usually assumed for TTSs. The situation improves when collisions are taken into account, as we will see in Sect.~\ref{sec:Discussion}. But our primary goal in this section, as already mentioned, is to present the method for solving the fluorescent line formation in the case $\epsilon = 0$, not to compare the results with observations.

\begin{table}[h!] \caption{Computation parameters for the outflow of Sect.~\ref{sec:accelerated-outflow}.} \label{tab:table1}
                \begin{tabular}{lllllllll}
                        \toprule
                        R$_c$& R$_{env}$ & M$_* $ &     \teff & \tenv &  $n_t(R_c)$ & $v_\infty$ & $v_{min}$ &   $\alpha$ \\
                        \midrule
                        2 R$_\sun$ & 10 R$_c$ & 1 M$_\sun$ & 4.5 $\cdot 10^3$K & $ 6 \cdot 10^{3}$ K & $2 \cdot 10^{16}$ cm$^{-3}$ & 250 km/s  & 0.02 & 0.95 \\
                        \bottomrule
                \end{tabular}
\end{table}

\subsection{Sobolev source functions\label{subsec:Sobolev-source-function}}

We consider three spectral transitions: $B,$ $C,$ and $R$ with rest wavelengths $\lambda{}_{B}$ \textless{} $\lambda{}_{C}$ \textless{}$\lambda{}_{R}$ in a wavelength interval 
\begin{equation}
        (\lambda_{R}-\lambda_{B})=\lambda_{R} \cdot v(R_{env})/c.
\end{equation}
For convenience, the lines are denoted $B$ for blue, $C$ for center,and $R$ for red; these lines may stem from different chemical elements and have respective thermal widths $\Delta\nu_{B}$, $\Delta\nu_{C}$, and $\Delta\nu_{R}$. 

The generalized Sobolev source functions of interacting lines can be computed by modifying the formalism developed by \citet{1978ApJ...219..654R} for the study of single lines emitted in velocity fields that lead to nonlocal radiative coupling. We assume in the following that readers are familiar with their work.

\subsubsection{$B$ line}

\begin{figure*}[tp]
        \centering
        \includegraphics[width=1.0\linewidth]{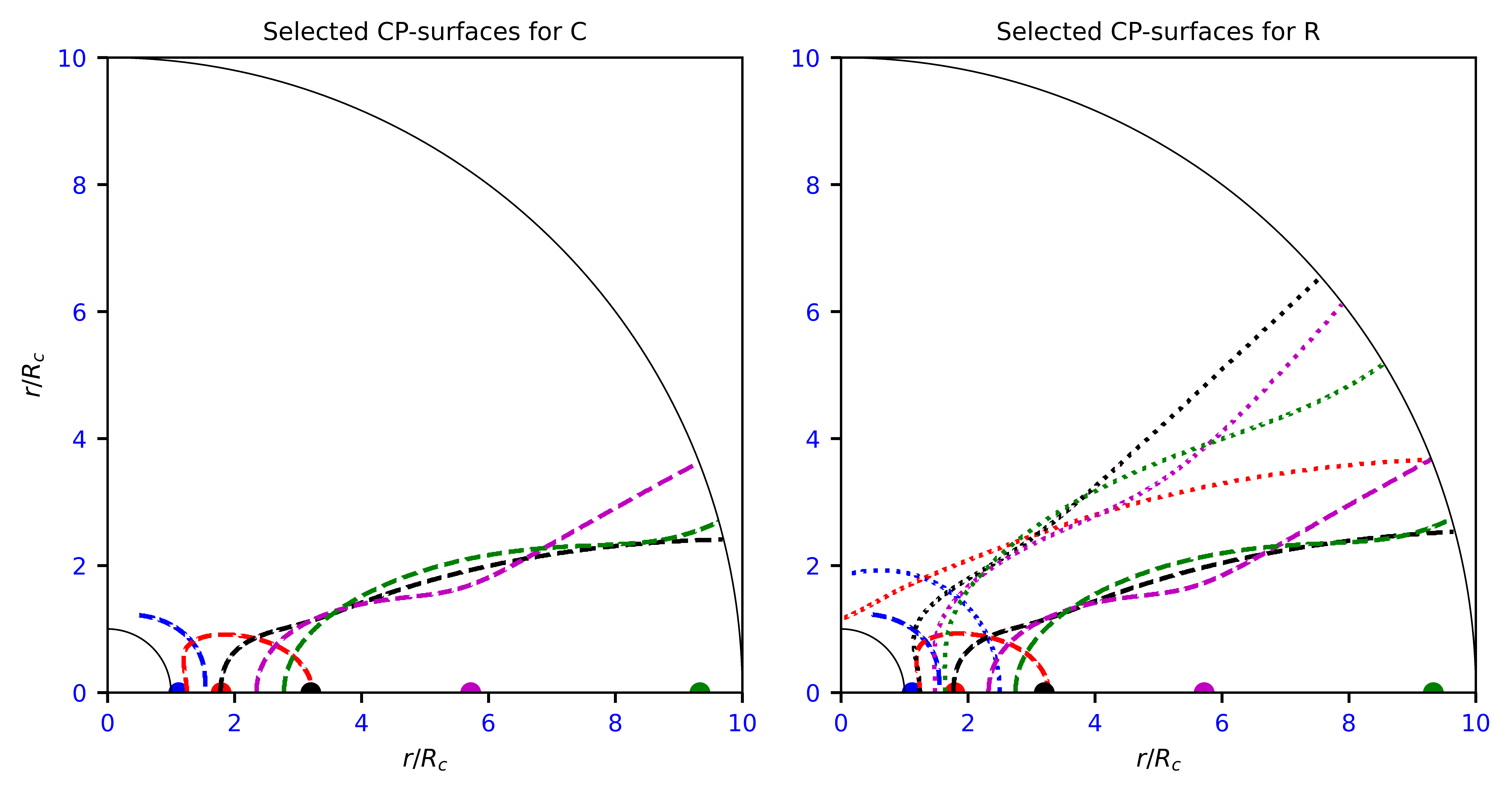}
        \caption[Short caption]{Selected CP surfaces for the spherically symmetric stellar wind, with the properties given in the text. Each surface is associated with the radial point of the same color. Note the variety of geometries for the surfaces, which give rise to discontinuities in the source function}
        \label{fig:figure2.3}
\end{figure*}

We first note that in any spherical, monotonically accelerating large-scale flow, photons emitted by the bluest line (the $B$ line as defined above) appear redshifted to atoms responsible for the redder $C$ and $R$ transitions and thus escape the medium freely. The line source of $B$ is therefore unaffected by $C$ and/or $R$, and its mathematical expression is the same as the one derived for a single line by Rybicki and Hummer. For an equivalent two-level atom, we then have 
\begin{equation}
S_{B}(r)=\frac{(1-\varepsilon_{B})\beta_{B}^{c}I_{B}^{c}+\varepsilon_{B}B_{B}}{\varepsilon_{B}+(1-\varepsilon_{B})\beta_{B}} \label{eq:SB}
,\end{equation}
where $\beta_{B}$ is the local escape probability, $\beta_{B}^{c}$ the probability that an escaping photon will strike the stellar core, $I_{B}^{c}$ is the background continuous radiation emitted by the stellar core at line frequency, $\varepsilon_{B}$ the ratio of collisional to total de-excitation rates from the upper level, and $B_{B}$ is the Planck function at rest line frequency and local electron temperature. All these quantities are defined locally at radius $r$. With $\mu=\cos\theta$ defining the direction cosine of a ray with respect to the outward radial direction, the local escape probability is given by 
\begin{equation}
\beta_{B}=\frac{1}{2}\intop_{-1}^{+1}\frac{1-e^{-\tau_{B}}}{\tau_{B}}d\mu,
\end{equation}
while the probability for an escaping photon to reach the stellar core is
\begin{equation}
\beta_{B}^{c}=\frac{1}{2}\intop_{\mu_{c}}^{+1}\frac{1-e^{-\tau_{B}}}{\tau_{B}}d\mu,
\end{equation}
where $\mu_{c}$ is the cosine of the half-angle subtended by the stellar core. The quantity $\tau_{B}$ in the last two equations is the $B$ line-center optical depth in direction $\mu$. Supposing that the transition occurs between electronic levels $l$ and $u$, $\tau_B$ is given by  \citep[see, e.g., ][Chapter 19, p.746ff] {2014tsa..book.....H}\begin{equation}
\tau_B(r, \mu) = \chi_{lu}(r) / Q(r, \mu) \label{Eq.tau} 
,\end{equation}
with the absorption coefficient given by
\begin{equation}
\chi_{lu}(r) =  (\pi e^2/mc)f_{lu}[n_l(r) - (g_l/g_u)n_u(r)] / \Delta \nu_D \label{Eq.chi}
\end{equation} 
and
\begin{equation}
Q(r, \mu) = |dv_s/ds| \label{Eq.Q}
,\end{equation}
where $v_s$ is the local fluid velocity projected in direction $s$ with cosine angle $\mu$, $\Delta \nu_D = \nu_0 v_{th}/c$ the local thermal line width, and where the other symbols have their usual meaning. The absorption coefficients of the three considered lines, as expressed by Eq.~\ref{Eq.chi}, are shown in Fig.~\ref{fig:figure2.2} for the parameters of Table \ref{tab:table1}.

\subsubsection{$C$ line}
We next consider photons of transition $C$ emitted in the outer parts of the envelope. They will interact with $B$ photons emerging from deeper parts of the envelope, where the emitting matter is moving slower, whenever the line-of-sight velocity difference is equal to the rest velocity difference between the two lines. In other words, interaction between $B$ and $C$ photons on any given line of sight will take place whenever 
\begin{equation}
v_{rad}(r_{B})-v_{rad}(r_{C})=\Delta\lambda_{BC}/c,\label{eq:CP_eq1}
\end{equation}                                     
where $\Delta\lambda_{BC}$ is the wavelength difference between lines $B$ and $C$. Solving Eq. \ref{eq:CP_eq1} over the entire space defines a locus of distant points $r^{'}$ where emitted $B$ photons can interact with $C$ photons emitted at $r$. This locus is called a common-point (CP) surface by \citet{1982ApJ...255..267O}, using the nomenclature introduced by \citet{1978ApJ...219..654R} to describe the  line self-coupling that occurs in those velocity laws that lead to distant radiative coupling. We use the same denomination and denote the CP surface described above by $r^{'}_{B}$ in the following equations. 

Figure~\ref{fig:figure2.3} shows some CP surfaces for the lines of interest in the stellar wind with properties shown in Fig.~\ref{fig:figure2.1} and parameters given in Table~\ref{tab:table1}. In that figure, the CP surfaces corresponding to a given radial point (identified by a colored dot in the figure) are represented by dashed and dotted lines sharing the same color as the radial point they are associated with. As mentioned earlier, the bluest line $\lambda$3968 is formed locally and has no associated CP surfaces. The dashed lines show the CP surfaces caused by interactions of the two redder lines $C$ and $R$ with their closest neighboring line (i.e., the interaction of $\lambda$3969 with $\lambda$3968 photons in the left panel, and interaction of $\lambda$3970 with $\lambda$3969 photons in the right panel of Fig.~\ref{fig:figure2.3}). Finally, the dotted lines denote the CP surface caused by interaction with the farthest spectral line (i.e., by the interaction of $\lambda$3970 with $\lambda$3968 photons; right panel). In other words, the left panel of Fig.~\ref{fig:figure2.3}, with only one set of CP surfaces, represents the situation for the $C$ line discussed here, while the right panel represents the situation for the $R$ line discussed in the next section. \footnote{Please note that we keep the same graphical conventions for the representation of CP surfaces in subsequent figures.} 

As mentioned by \citet{1982ApJ...255..267O} and also seen in our figures, the topology of CP surfaces is quite complex and varies with radial distance from the stellar core. Close to the star, the CP surface is an arc of circle surrounding the reference point and extending from the abscissa to the angle at which the line of sight intercepts the core. At larger distances it becomes a closed surface entirely surrounding the reference point, which then opens farther out as the envelope limit is reached. We see below that these topological changes are reflected in the source functions of optically thick lines.

The source function for transition $C$ takes into account the presence of resonant CP surfaces coupling it to the $B$ transition. It is written as (see also Olson 1982)
\begin{equation}
S_{C}(r)=\frac{(1-\varepsilon_{C})\beta_{C}^{c}I_{C}^{c}+\varepsilon_{C}B_{C}+(1-\varepsilon_{C})S_{nl}^{C}(r)}{\varepsilon_{C}+(1-\varepsilon_{C})\beta_{C}},\label{eq:SC}
\end{equation}
where the nonlocal contribution $S_{nl}^{C}$ to the source is 
\begin{equation}
S^{nl}_{C}=\frac{1}{2}\intop_{-1}^{+1}\frac{1-e^{-\tau_{C}}}{\tau_{C}}S(r^{'}_{B})(1-e^{-\tau(r^{'}_{B})})d\mu\label{eq:SCnl}
\end{equation}
and the probability that a photon strikes the core is modified to take into account the intervening $B$ layer: 
\begin{equation}
\beta_{C}^{c}=\frac{1}{2}\intop_{\mu_{c}}^{+1}\frac{1-e^{-\tau_{C}}}{\tau_{C}}e^{-\tau(r^{'}_{B})}d\mu.\label{eq:BetaCc}
\end{equation}

\subsubsection{$R$ line}
Similarly, the third and most redward transition $R$ will be affected by photons of both lines $C$ and $B$, and we must locate the two CP surfaces $r^{''}_{B}$ and $r^{'''}_{C}$ where $B$ and $C$ have the same line-of-sight velocity as the gas at the considered $R$ emission point $r$, using Eq.~\ref{eq:CP_eq2}:
\begin{equation} \left\{
\begin{array}{l}
                v_{rad}(r_{R})-v_{rad}(r_{B})=\Delta\lambda_{BR}/c \\
                v_{rad}(r_{R})-v_{rad}(r_{C})=\Delta\lambda_{RC}/c \label{eq:CP_eq2}
\end{array} \right. \
.\end{equation}
The generalization of Eqs. \ref{eq:CP_eq1} to \ref{eq:BetaCc} to line $R$ is obvious and leads to the following expression for $S_{R}(r)$:
\begin{equation}
        S_{R}(r)=\frac{(1-\varepsilon_{R})\beta_{R}^{c}I_{R}^{c}+\varepsilon_{R}B_{R}+(1-\varepsilon_{R})S_{nl}^{R}(r)}{\varepsilon_{R}+(1-\varepsilon_{R})\beta_{R}},\label{eq:SR}
\end{equation}
with the nonlocal integral $S_{R}^{nl}$ given by
\begin{equation}
S^{nl}_{R}=\frac{1}{2}\intop_{-1}^{+1}\frac{1-e^{-\tau_{R}}}{\tau_{R}} \left [ S(r^{''}_{B})(1-e^{-\tau(r^{''}_{B})}) + S(r^{'''}_{C})(1-e^{-\tau(r^{'''}_{C}}))e^{-\tau(r^{''}_{B})} \right ] d\mu
\end{equation}
and 
\begin{equation}
\beta_{R}^{c}=\frac{1}{2}\intop_{\mu_{c}}^{+1}\frac{1-e^{-\tau_{R}}}{\tau_{R}}e^{-\tau(r^{''}_{B})-\tau(r^{'''}_{C})}d\mu.\label{eq:BetaRc}
\end{equation}

The geometry of CP surfaces, where the various quantities entering Eqs.~\ref{eq:SCnl} to~\ref{eq:BetaRc} are computed, depends only on the gas velocity law and the transitions wavelengths. It is computed numerically by integrating over (a) the entire space when computing the nonlocal parts of the source functions and (b) the angle subtended by the stellar core when computing the stellar continuum contribution to the source. The computation of $S_{B}$ is straightforward algebra in this approximation since all quantities are local, while computing $S_{C}$ and $S_{R}$ requires an iterative process to solve  for the nonlocal parts $S^{nl}$ consistently. 

\begin{figure}
        \centering
        \begin{minipage}{0.48\textwidth}
                \centering
                \includegraphics[width=0.95\textwidth]{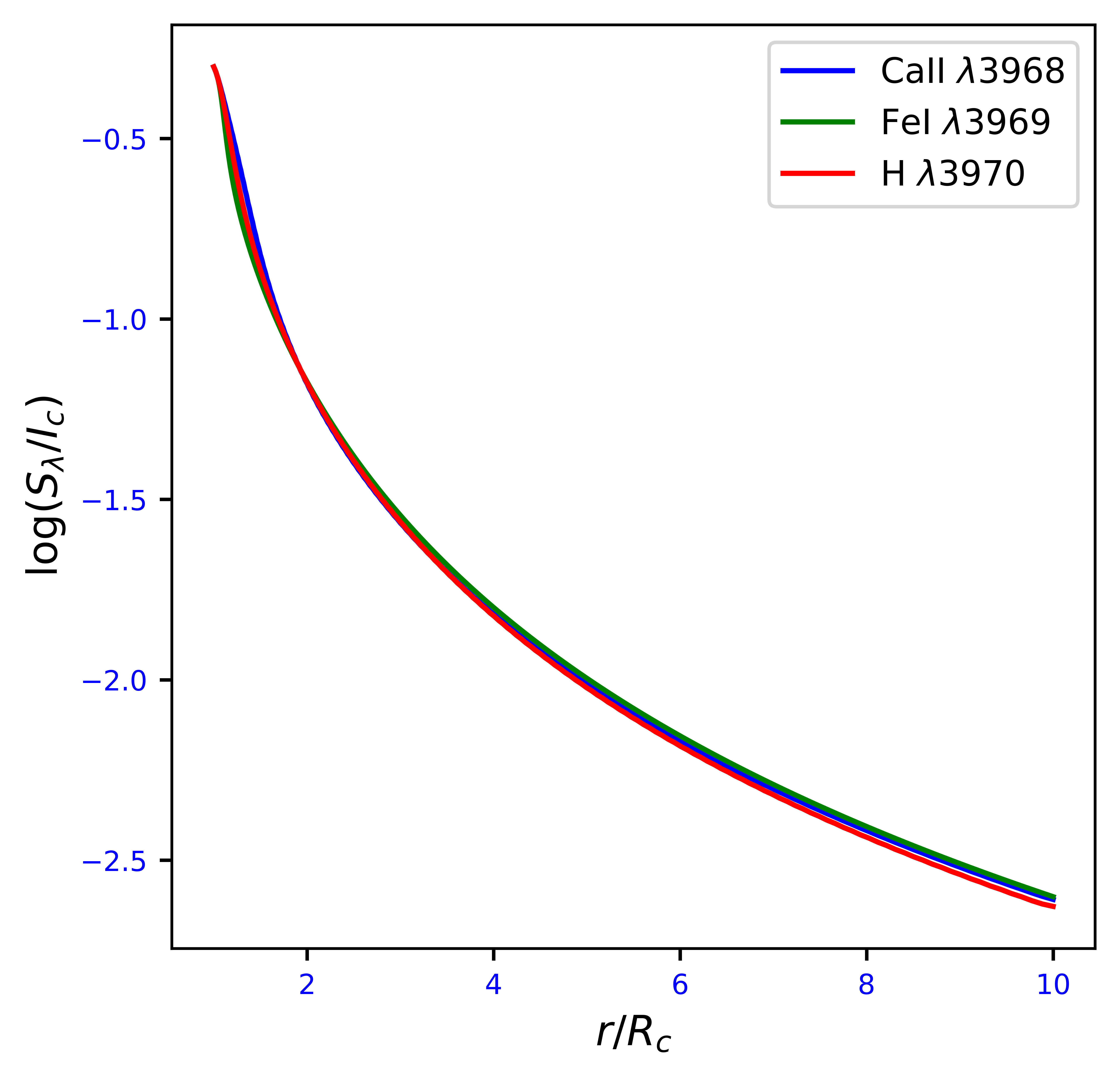} % first figure itself
                \caption[Short caption]{Source functions for the three lines when they are computed independently.}
                \label{fig:figure2.4}
        \end{minipage}\hfill
        \begin{minipage}{0.48\textwidth}
                \centering
                \includegraphics[width=0.95\textwidth]{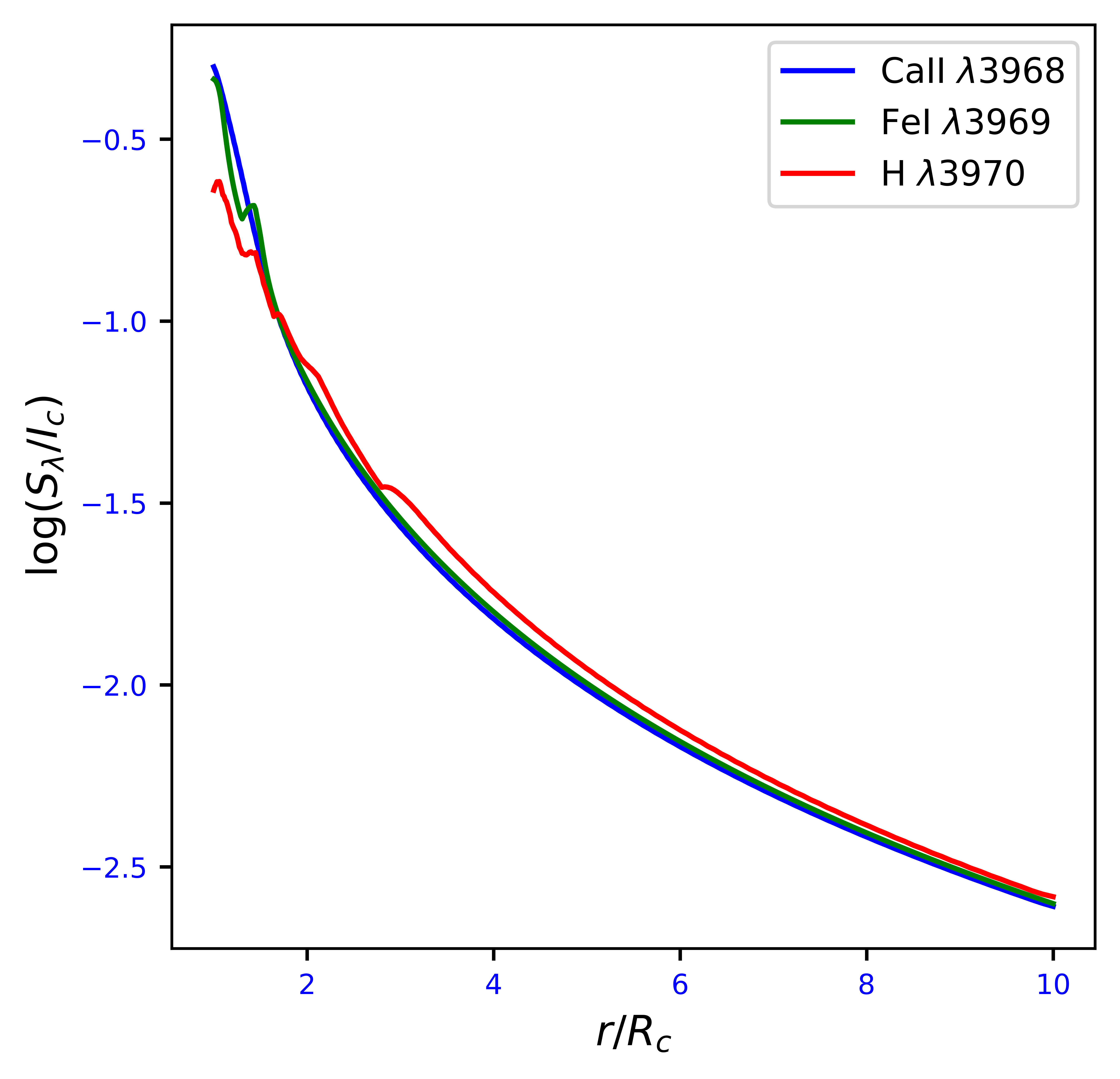} % second figure itself
                \caption[Short caption]{Source functions for the three lines when radiative interactions are taken into account.}
                \label{fig:figure2.5}
        \end{minipage}
\end{figure}

The above discussion suggests constructing an interaction matrix $I$ to keep track of each possible line interaction, a concept that proved useful in coding the problem at hand. The matrix elements are  0 for a line that can be considered single or 1 when an interaction is present. For the current case of three lines in an accelerating outflow, with line ordering $B, C, R$, the interaction matrix is
\begin{equation}
        I = \left(\begin{array}{ccc}
                0 & 0 & 0 \\ 1 & 0 & 0 \\ 1 & 1 & 0
        \end{array}\right),
\end{equation}
meaning that the $B$ line (first row) has no interaction with itself, $C$, or $R$, while the  $C$ line (second row) interacts with $B$, and the $R$ line (third row) interacts with both $B$ and $C$.

Figure~\ref{fig:figure2.4} displays the computed source functions for the three lines under consideration when they are treated as independent, that is to say, when local equations similar to Eq.~\ref{eq:SB} are used for computing all of them. Figure~\ref{fig:figure2.5} shows the source functions computed when considering the radiative interactions (i.e., when using Eq.~\ref{eq:SC} for the $C$ line and Eq.~\ref{eq:SR} for the $R$ line). While the source function of the $B$ line is evidently the same in both cases, discontinuities caused by the different topologies of CP surfaces are seen; but since the lines are only moderately optically thick, the induced changes remain small and do not affect the resulting source functions very much. The assumption of LTE atomic populations, which results in constant radial optical depths as the source functions are iterated, also plays a role here. Allowing for an iterative estimate of populations is thus likely to result in very different source functions. While the two-level atom assumption is clearly a limitation of the current code, it is a first step that allows us to understand the radiative couplings of the lines occurring in both the comoving reference frame and the observer's frame, and the next section deals with the latter.

\subsection{Exact line profile integration\label{sec:Exact-line-profile}}

\begin{figure*}[tp]
        \centering
        \includegraphics[width=1.0\linewidth]{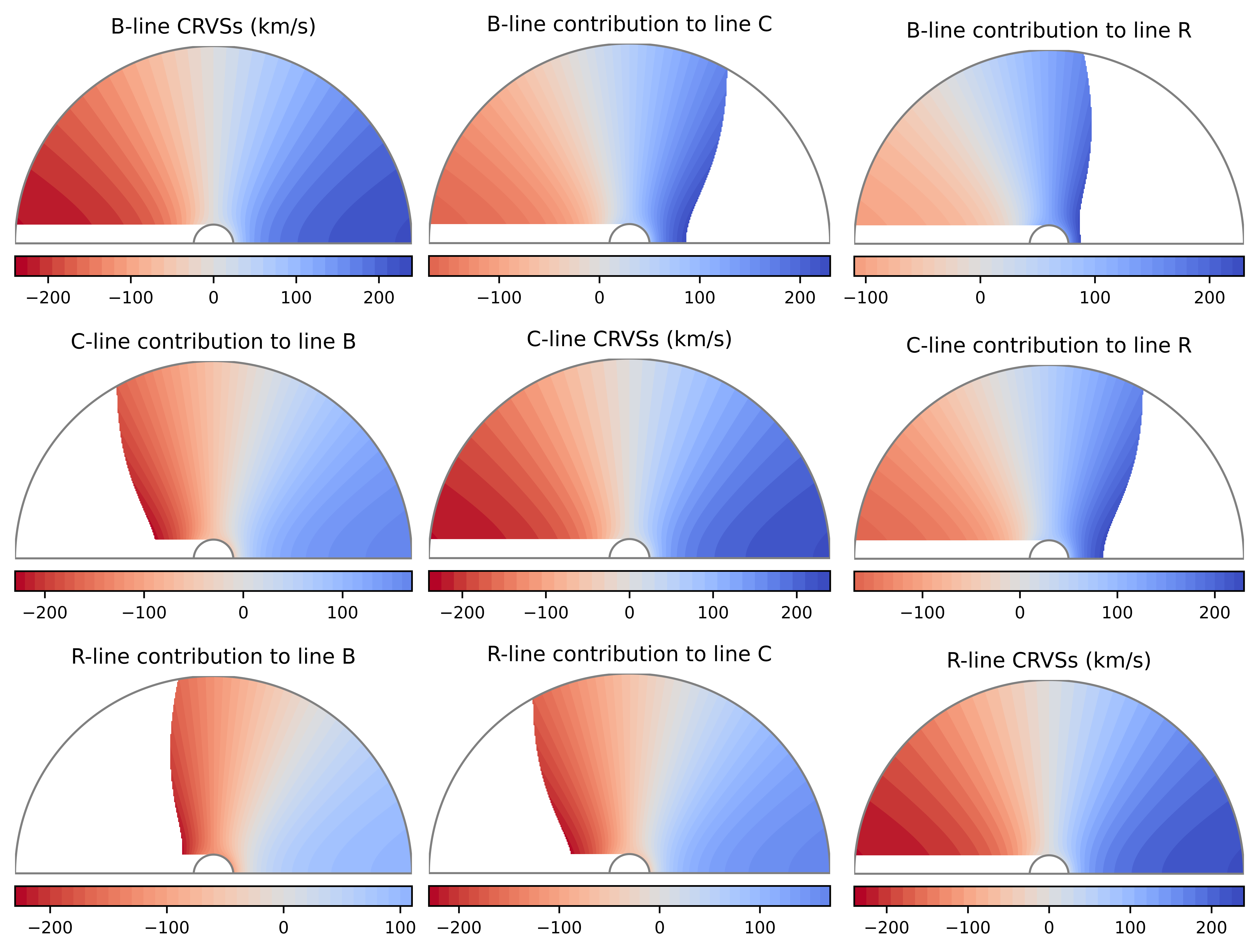}
        \caption[]{Regions and frequencies contributing to the line profiles of  the $B$, $C$, and $R$ lines (respectively for each column) in the wind model of Sect. 2. Shown here is a meridional cut in the half spherical stellar envelope of radius $R_{max} = 10 R_c$ surrounding a stellar core of radius $R_c$. Displayed are the constant radial velocity surfaces (CRVSs) seen by an outside observer located at infinity on the right side of the figure for the three lines considered (left-right diagonal), as well as the displaced CRVSs of neighboring lines. The scale of the color bars is in km/s, and blue (red) indicates positive (negative) line frequencies. }
        \label{fig:figure2.6}
\end{figure*}

We refer to \citet{1984ApJ...285..269B} for a detailed discussion of the line profile integration method for single lines and focus here on the specific modifications that are needed to introduce the possibility of line interaction in the flux integration. The emergent flux $F_{x,l}$ of line~$l$ 
%(normalized to the underlying stellar continuum $F_{x,l}^{c}$) 
at frequency displacement $x=(\nu-\nu_{0,l})/\Delta\nu_{l}$ is 
\begin{equation}
F_{x,l}=2\pi \int_{0}^{R_{max}}I_{l}(p,x,z=\infty)pdp,
\end{equation}
where impact parameter $p$ and abscissa $z$ are related to the usual $r$ and $\theta$ coordinates by the relationships $r^{2}=p^{2}+z^{2}$ and $z=r\cos\theta$. The quantity $I_{l}(p,x,z=\infty)$ is the emergent intensity at frequency $x$ on the ray $p$, and $\Delta\nu_{l}$ is the width of the local absorption profile. The Doppler equation at $(p,z)$ is 
\begin{equation}
x(p,z)=\frac{v(r)}{\xi_{l}(r)}\frac{z}{r},
\end{equation}
where $\xi_{l}(r)$ is the width of the line absorption profile in the same units as $v(r)$. The local absorption profile $\xi_{l}(r)$ results from both thermal broadening and an additional turbulent widening $v_{turb}$, chosen to be equal to 1km/s in the actual computations. The only constraint on the choice of the overall absorption profile width in velocity space is that its integral over velocity (or frequency) be the same for the three considered lines (see below). To provide a frequency rest frame common to these lines, we thus chose the width $\xi$ to be equal to or larger than the widest local line width expressed in velocity units. In other words, 
\begin{equation}
\xi\equiv(\alpha+1){\rm max}[\xi_{l}(r)],
\end{equation}
where $\alpha$ is typically in the range $[0, 1]$.

In an accelerated outflow, there is a unique correspondence between line frequencies and positions on the line of sight. This is shown in the top left panel of Fig.~\ref{fig:figure2.6}, which displays the envelope's gas radial velocities seen by an outside observer located at $z = \infty$. The color code of radial velocities is given in the bar under the figure. Impact parameter $p=0$ contains the full range of velocities (or, equivalently, frequency displacements) present in the three emergent line profiles, ranging from $-V_{max}$ at $z=-R_{max}$ to $+V_{max}$ at $z=R_{max.}$ We set up a uniform grid over this velocity range with the understanding that the velocity step $dv$, which defines the numerical resolution of the line profile computation, is small enough to allow the narrowest line absorption profile, which has width $\xi_{l}$, to be resolved with sufficient accuracy. If we consider an impact parameter~$p_{j}$, it will contain velocities in the range 
\begin{equation}
v_{rad}^{min}=V_{max}\frac{z_{min}}{R_{max}}\leq v_{rad}\leq V_{max}\frac{z_{max}}{R_{max}}=v_{rad}^{max}
\end{equation}{\tiny }
with 
\[
z_{min}=\Biggr\{\begin{array}{cc}
\sqrt{R_{c}^{2}-p_{j}^{2}}, & p_{j}\leq R_{c}\\
-\sqrt{R_{max}^{2}-p_{j}^{2}}, & p_{j}>R_{c}
\end{array}{\rm {\rm \;and}\quad}z_{max}=\sqrt{R_{max}^{2}-p_{j}^{2}}.
\]

\subsubsection{Line-$C$ flux}

In the observer's frame, there is no preferred flow direction, so all transitions are treated in the same way. We now consider the formation of line $C$ in some detail, since it is our primary candidate for fluorescence, with the understanding that the two other lines are formed in a similar way. For the following discussion, the reader is invited to refer to the central column of Fig.~\ref{fig:figure2.6}, which illustrates the origins of line $C$.

When they reach the observer at $z = \infty$, line-$C$ photons with apparent frequency $\tilde{x}$ traveling in the $p$~direction originate in atomic transitions taking place in three different regions;
\begin{enumerate}
        \item [(a)] Transition $C$ photons are emitted from atoms in the $z$-range corresponding to frequencies around $\tilde{x}$ in the local line emission profile $\Delta \nu_{D_{C}}$. This is the situation illustrated by the central figure of Fig.~\ref{fig:figure2.6}, which displays the constant radial velocity surfaces (CRVSs) of $C$-emitting atoms seen by the observer. As explained above, each of the CRVSs corresponds to a frequency $\tilde{x}$ in the overall line profile. 
        \item [(b)] A second contribution to the $C$-line intensity on $p$ stems from transition $B$ photons that are emitted by atoms moving slower by the velocity difference between the two lines $c \Delta \nu_{BC} / \nu_{C}  \pm \Delta \nu_{D_{C}}$ and are therefore located deeper in the envelope (as seen by the observer) than atoms emitting transition $C$ frequency $\tilde{x}$. The top panel of the central column in Fig.~\ref{fig:figure2.6} displays the frequency range of this contribution or, equivalently, the envelope parts from which it originates.
        \item [(c)] Finally, there is a third contribution from transition $R$ photons emitted by atoms that are moving faster by $c \Delta \nu_{CR} / \nu_{C} \pm \Delta \nu_{D_{C}}$  and are therefore located  in front of the atoms that emit transition $C$ at frequency $\tilde{x}$. The bottom panel of the central column in Fig.~\ref{fig:figure2.6} displays the space-frequency distribution of this contribution.
\end{enumerate}
 Similar considerations hold for the other two lines and the resulting space-frequency contributions to these lines are illustrated by the first and third columns of Fig.~\ref{fig:figure2.6}. 

We next considered the contributions (a) to (c) above to the optical depth and emergent $C$-line intensity at $\tilde{x}$  in a more quantitative way. 

\paragraph{Local contribution to the C-line intensity.}

The $z$-range corresponding to the local absorption profile frequency range is $z(\tilde{x}-\xi/2)<\tilde{z}<z(\tilde{x}+\xi/2)$, and the optical depth along $p$ is given by 
\begin{equation}
\tau_{C}(p,z,\tilde{x})=\frac{1}{\sqrt{\pi}}\int_{z(\tilde{x}-\xi/2)}^{z}\frac{\kappa_{C}(r')}{\Delta\nu_{D_{C}}(r')}\varphi(x')dz',\label{eq:tau_Cc}
\end{equation}
where $\kappa_{C}$ is the line center opacity and $\varphi$ the line absorption profile (see also Eqs.~\ref{Eq.tau} to \ref{Eq.chi}). 
 Let us denote with $\tau_C^*$ the maximum value\footnote{We use a similar notation for the maximum values of the other lines' optical depths.} of $\tau_C$ on the ray: 
\begin{equation}
        \tau_{C}^*=\frac{1}{\sqrt{\pi}}\int_{z(\tilde{x}-\xi/2)}^{z(\tilde{x}+\xi/2)}\frac{\kappa_{C}(r)}{\Delta\nu_{D_{C}}(r)}\varphi(x)dz.
\end{equation}
 Moving in the $-z$ direction along impact parameter $p$,  and remembering that the line frequency displacement decreases with $z$ until the line center is reached at $z = 0$, the line optical depth then varies as
\begin{equation}
\tau_{C}(p,z,\tilde{x})=\left\{
\begin{aligned}
        0, & \quad z > z(\tilde{x}+\xi/2)\\
        \frac{1}{\sqrt{\pi}}\int_{z(\tilde{x}+\xi/2)}^{z}\frac{\kappa_{C}(r')}{\Delta\nu_{D_{C}}(r')}\varphi(x')dz', & \quad  z(\tilde{x}+\xi/2) \geq z >  z(\tilde{x}-\xi/2)\\
        \tau_C^*, & \quad  z <  z(\tilde{x}-\xi/2).\\
        \end{aligned} \right.
\end{equation}
In other words, the optical depth varies from $0$ to $\tau_C^*$ over a short distance and then remains constant over the ray. The line intensity at $\tilde{x}$ is 
\begin{equation}
        I_{C}^{c}(p,z,\tilde{x})=\int_{0}^{\tau(\tilde{x})}S_{C}(p',z',\tilde{x}')\exp[-\tau_{C}(p',z',\tilde{x}')]d\tau(p',z',\tilde{x}'),\label{eq:I_Cc}
\end{equation}
where the $c$ subscript indicates that this intensity part is caused by $C$-line photons.

Following \citet{1984ApJ...285..269B}, we rewrite Eq.~\ref{eq:tau_Cc} as 
\begin{equation}
\tau_{C}(p,z,x_{j})=\sum_{ij}\Delta\tau_{ij}\label{eq:tau_sig}
\end{equation}
with 
\begin{equation}
\Delta\tau_{ij}=\frac{1}{\sqrt{\pi}}\biggl\langle\frac{\kappa_{l,i}}{\Delta\nu_{D_{C}}}\left( \dfrac{dx}{dz}\right )^{-1}_{i}\biggr\rangle\int_{x_{j-1/2}}^{x_{j+1/2}}\varphi(x')dx'.\label{eq:del_tau}
\end{equation}

In Eq.~\ref{eq:tau_sig}, the summation over $i$ runs over the points $z_{i}$ of the $z$ grid defined on impact parameter $p,$ while the summation over j runs over frequency displacements at which a contribution is expected at $z_{i}$ (i.e., for $x_{j}$ in the range
\noindent $x_{i}-\xi/2<x_{j}<x_{i}+\xi/2$, where $x_{i}=x(z_{i})$).
The averaged quantity in Eq. \ref{eq:del_tau} is evaluated over the frequency step $\delta x$, and the integral is equal to $[{\rm {erf}(x_{j+1/2})-{\rm {erf}(x_{j-1/2})]/2}}$ if the local absorption profile is Gaussian.

Equation~\ref{eq:I_Cc} can now be integrated to obtain the emergent intensity of line $C$ if it were a single line without interactions with its neighbors:
\begin{equation}
I_{C}^{c}(p,z=\infty,\tilde{x})=I_{C}^{*}(\tilde{x})e^{-\tau_{C}}+S_{C}(1-e^{-\tau_{C}})\label{eq:integrated_I_Cc}
,\end{equation}
where $I_{C}^{*}$ is the stellar continuum intensity at $\lambda_{C}$ and $\tau_{C}$ the optical depth value at $\tilde{x}$ on $p$, given by Eqs.~\ref{eq:tau_sig} and~\ref{eq:del_tau}. 

While the similarity with the expression derived in the framework of the Sobolev theory is evident, we did not factor out the profile function, but integrated it explicitly over the region where the optical depth goes from zero to its maximum value, $\tau_C^*$, on the ray. In other words, we did not make use of the Sobolev approximation, so that Eq.~\ref{eq:integrated_I_Cc} is equivalent to Eq.~25 in \citet{1978ApJ...219..654R} rather than to Eq.~26 of the same paper. 

We next consider the additional contributions to the intensity on $p$ caused by the neighboring lines.  

\begin{figure}
        \centering
        \begin{minipage}{0.48\textwidth}
                \centering              \includegraphics[width=0.95\textwidth]{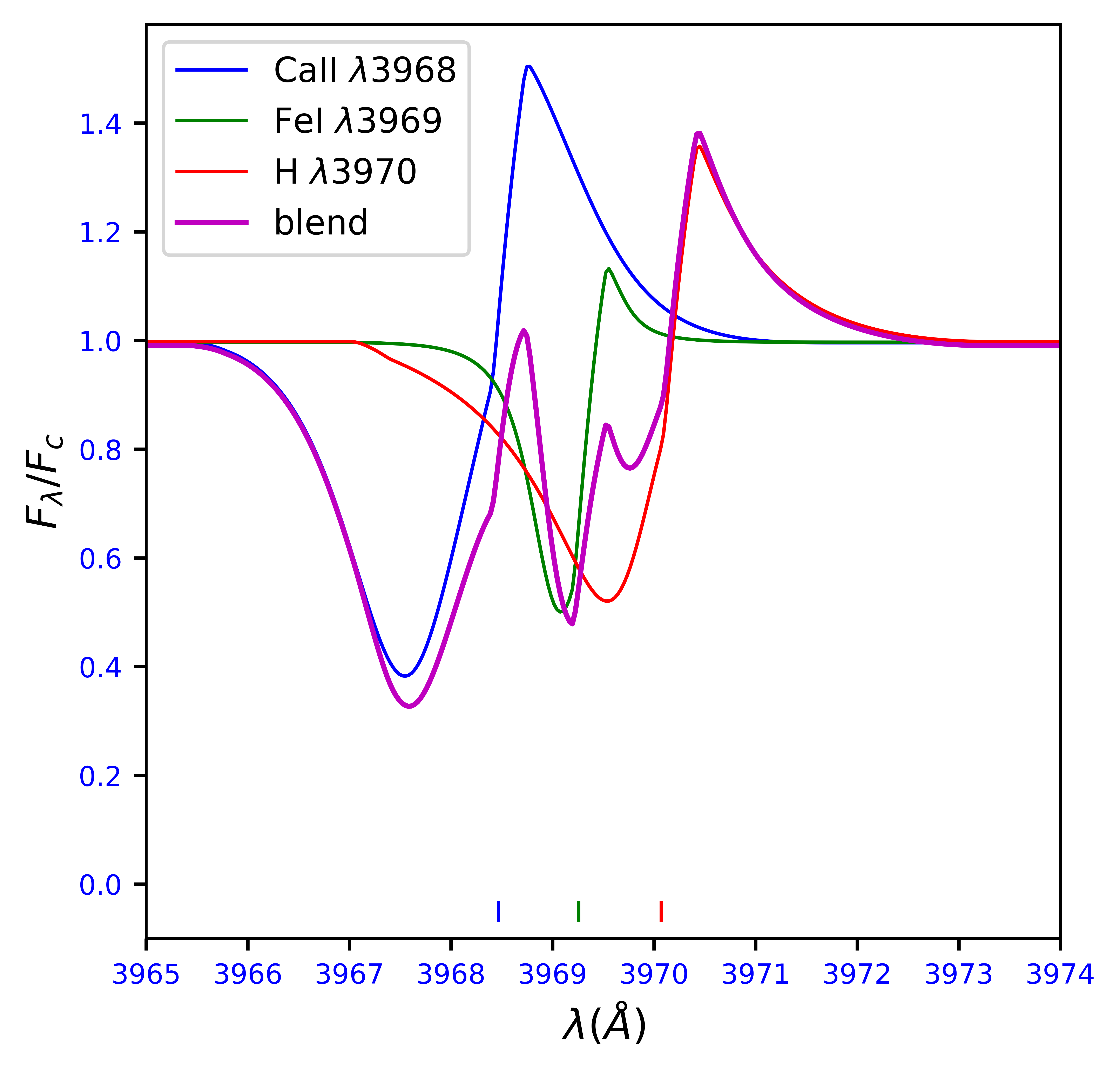} % first figure itself
                \caption[Short caption]{Line profiles of the three lines considered and their blend for the wind model of Sect. 2 when the lines are computed independently.}
                \label{fig:figure2.7}
        \end{minipage}\hfill
        \begin{minipage}{0.48\textwidth}
                \centering
                \includegraphics[width=0.95\textwidth]{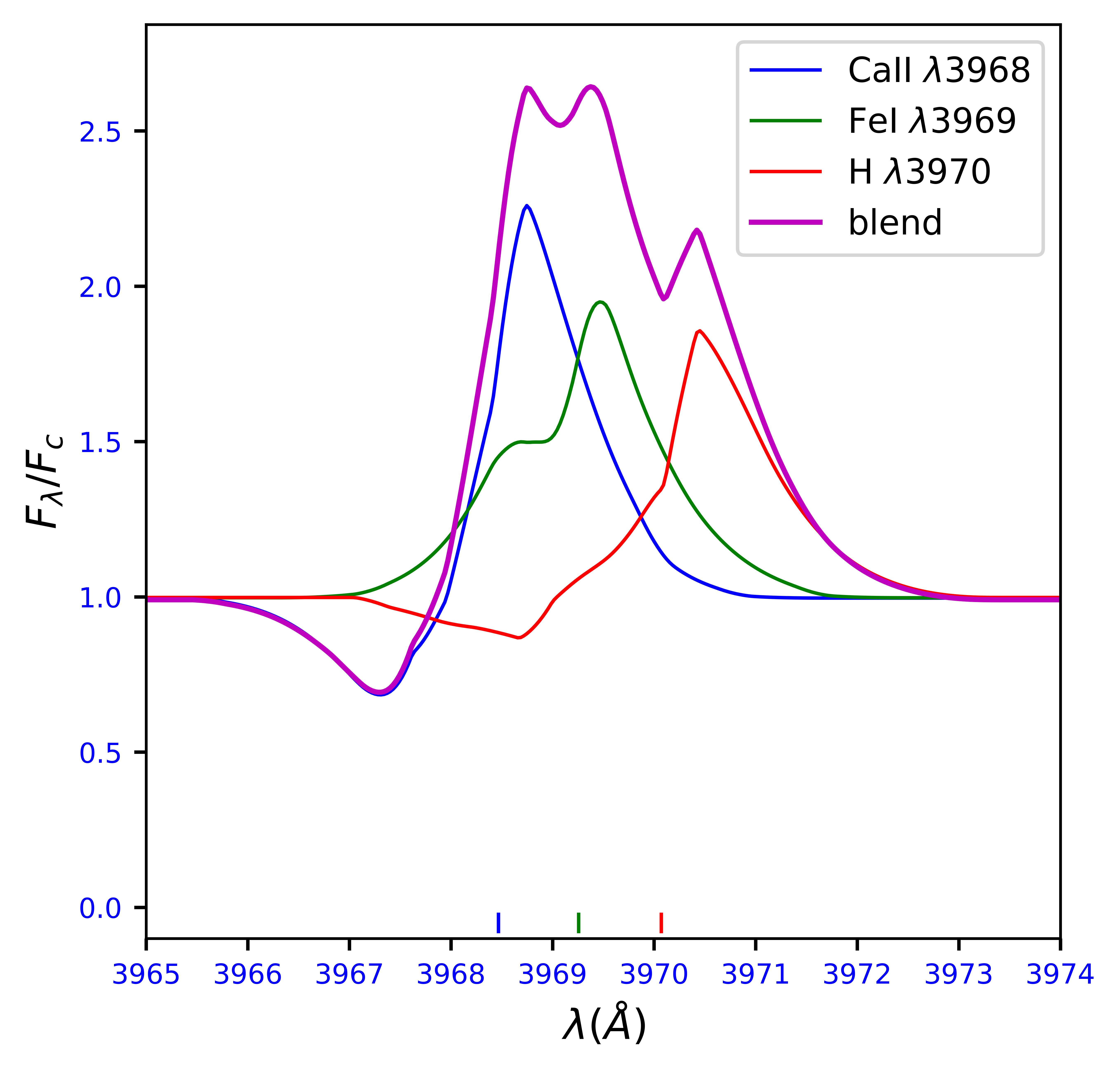} % second figure itself
                \caption[Short caption]{Line profiles of the three lines considered and their blend for the wind model of Sect. 2 when radiative interactions between lines are taken into account.}
                \label{fig:figure2.8}
        \end{minipage}
\end{figure}

\paragraph{Contribution from B-line photons to line C.}

As noted earlier, line~$B$ is displaced by $\delta v$ from line~$C$ such that line-$B$ photons traveling in the $z$ direction on a given impact parameter will be able to interact with atoms emitting transition~$C$ that are located at apparent velocities
\begin{equation}
v_{rad}^{min}\leq v_{rad}\leq v_{rad}^{max}-\delta v
\end{equation}
since these are Doppler-shifted to the rest frequency of line~$C$. As illustrated by the top central panel of Fig.~\ref{fig:figure2.6}, line~$C$ will get a contribution from~$B$ photons that are emitted deeper in the envelope and are moving slower by an amount $\delta v$. This contribution is taken into account by additional terms in Eqs.~\ref{eq:tau_Cc} and~\ref{eq:I_Cc}. Naming $\delta x$ the frequency displacement corresponding to $\delta v$, the optical depth of line $C$ along the line of sight is augmented by the quantity
\begin{equation}
        \tau_{B}(p,z,\tilde{x})=\frac{1}{\sqrt{\pi}}\int_{z(\tilde{x}-\xi/2-\delta x)}^{z}\frac{\kappa_{B}(r')}{\Delta\nu_{D_{B}}(r')}\varphi(x')dz',
\end{equation}
while the line intensity becomes
\begin{equation}
        I_{C}^{bc}(p,z,\tilde{x}) = \int_{0}^{\tau(\tilde{x})} \{ S_{C}(p',z',\tilde{x}')\exp[-\tau_{C}(p',z',\tilde{x}')] + S_{B}(p',z',\tilde{x}') \exp[-\tau_{B}(p',z',\tilde{x}')-\tau_{C}(p',z',\tilde{x}')] \} d\tau(p',z',\tilde{x}') \label{eq:I_Cbc}
\end{equation}
since line $B$ forms deeper than $C$ (as seen by an outside observer). The subscript $bc$ indicates that the expression for the $C$ intensity now takes contributions of both $B$ and $C$ into account. The optical depth along impact parameter $p$ for frequency $\tilde{x}$, from $z =z_{max}(p)$ inward, varies as\footnote{This assumes that $p$ is close to the stellar core and $\tilde{x}$ is both positive and lies in the vicinity of line center, so that the entire frequency range of interest is present on the impact parameter.}
\begin{equation}
\tau_{C}^{bc}(p,z,\tilde{x})=\left\{
        \begin{aligned}
0,& \quad  z > z(\tilde{x}+\xi/2)\\
\frac{1}{\sqrt{\pi}}\int_{z(\tilde{x}+\xi/2)}^{z}\frac{\kappa_{C}(r')}{\Delta\nu_{D_{C}}(r')}\varphi(x')dz',  & \quad  z(\tilde{x}+\xi/2 \geq z > z(\tilde{x}-\xi/2)\\
\tau_C^*,  & \quad z(\tilde{x}-\xi/2) \geq  z > z(\tilde{x}+\xi/2 - \delta x)\\
\tau_C^*+  \frac{1}{\sqrt{\pi}}\int_{z(\tilde{x}+\xi/2-\delta x)}^{z}\frac{\kappa_{B}(r')}{\Delta\nu_{D_{B}}(r')}\varphi(x')dz', & \quad z(\tilde{x}+\xi/2-\delta x) > z \geq z(\tilde{x}-\xi/2-\delta x)\\
\tau_C^*+ \tau_B^* , & \quad z < z(\tilde{x}-\xi/2 - \delta x).
        \end{aligned} \right.
\end{equation}

\noindent We then integrate Eq.~\ref{eq:I_Cbc} to obtain the emergent intensity at $\tilde{x}$:
\begin{equation}
I_{C}^{bc}(p,z=\infty,\tilde{x})=I_{C}^{*}(\tilde{x})e^{-\tau_{B}-\tau_{C}}+S_{C}(1-e^{-\tau_{C}})+S_{B}(1-e^{-\tau_{B}})e^{-\tau_{C}}\label{eq:integrated_I_Cbc},
\end{equation}
where the notations are as defined in the previous paragraphs. 

\paragraph{Contribution from R-line photons to line C.}

We next consider the contribution of the reddest of these three lines to $C$, which is displaced from $C$ by a velocity amount~$-\delta v^\dagger$ (or the corresponding frequency displacement $-\delta x^\dagger$), and interactions will occur with atoms moving at velocities
\begin{equation}
        v_{rad}^{min}\leq v_{rad}\leq v_{rad}^{max}+\delta v^\dagger.
\end{equation}
  While photons of line $B$ contributing to $C$ emerge from deeper layers than those of line $C$, we now have the opposite situation: line-$R$ photons contributing to $C$ originate from layers above $C$ (as seen by an observer at $z=\infty$). The optical depth along the line of sight is increased by an amount 
\begin{equation}
        \tau_{R}(p,z,\tilde{x})=\frac{1}{\sqrt{\pi}}\int_{z(\tilde{x}-\xi/2-\delta x^\dagger)}^{z}\frac{\kappa_{R}(r')}{\Delta\nu_{D_{R}}(r')}\varphi(x')dz',
\end{equation}
  while the line intensity becomes
\begin{equation}
\begin{split}
  I_{C}^{bcr}(p,z,\tilde{x}) = &  \int_{0}^{\tau(\tilde{x})} \{ S_{B}(p',z',\tilde{x}')\exp[-\tau_{B}(p',z',\tilde{x}')-\tau_{C}(p',z',\tilde{x}')-\tau_{R}(p',z',\tilde{x}')] \\
  &     + S_{C}(p',z',\tilde{x}') \exp[-\tau_{C}(p',z',\tilde{x}')-\tau_{R}(p',z',\tilde{x}')] + S_{R}(p',z',\tilde{x}') \exp[-\tau_{R}(p',z',\tilde{x}')] \} d\tau(p',z',\tilde{x}'), \label{eq:I_Cbcr}
\end{split}
\end{equation}
where the subscript $bcr$ indicates that contributions of the three interacting lines to $C$ are now taken in account. The optical depth variation along the ray is written, with the same caveat as before, as
\begin{equation}
        \tau_{C}^{bcr}(p,z,\tilde{x})=\left \{
        \begin{aligned}
                0, & \quad z > z(\tilde{x}+\xi/2+ \delta x^\dagger)\\
                \frac{1}{\sqrt{\pi}}\int_{z(\tilde{x}+\xi/2+\delta x^\dagger)}^{z}\frac{\kappa_{R}(r')}{\Delta\nu_{D_{R}}(r')}\varphi(x')dz',  & \quad  z(\tilde{x}+\xi/2- \delta x^\dagger)\geq z > z(\tilde{x}-\xi/2+\delta x^\dagger)\\
                \tau_R^*, &  \quad z(\tilde{x}-\xi/2+\delta x^\dagger) \geq  z > z(\tilde{x}+\xi/2)\\
                \tau_R^* +  \frac{1}{\sqrt{\pi}}\int_{z(\tilde{x}+\xi/2)}^{z}\frac{\kappa_{C}(r')}{\Delta\nu_{D_{C}}(r')}\varphi(x')dz',  &  \quad z(\tilde{x}+\xi/2) > z \geq z(\tilde{x}-\xi/2)\\
                \tau_R^*+ \tau_C^*, &  \quad  z(\tilde{x}+\xi/2) \geq z > z(\tilde{x}+\xi/2 - \delta x) \\
                \tau_R^* + \tau_C^* + \frac{1}{\sqrt{\pi}}\int_{z(\tilde{x}+\xi/2-\delta x)}^{z}\frac{\kappa_{B}(r')}{\Delta\nu_{D_{B}}(r')}\varphi(x')dz', & \quad z(\tilde{x}+\xi/2-\delta x ) > z \geq  z(\tilde{x}-\xi/2-\delta x )\\
                \tau_R^* + \tau_C^*+ \tau_B^*,  & \quad z < z(\tilde{x}-\xi/2 - \delta x).
        \end{aligned}
        \right.
\end{equation} 

\noindent Integrating Eq~\ref{eq:I_Cbcr}, we obtain the final emergent intensity of line $C$, which now takes all interactions into account:
\begin{equation}
I_{C}^{bcr}(p,z=\infty,\tilde{x})=I_{C}^{*}(\tilde{x})e^{-\tau_{B}-\tau_{C}-\tau_{R}}+S_{C}(1-e^{-\tau_{C}})e^{-\tau_{R}}+S_{B}(1-e^{-\tau_{B}})e^{-\tau_{C}-\tau_{R}}+S_{R}(1-e^{-\tau_{R}}). \label{eq:integrated_I_Cbcr}
\end{equation}
The line flux then follows from an integration of Eq.~\ref{eq:integrated_I_Cbcr} over all impact parameters.

\subsubsection{Lines $B$ and $R$}

As mentioned earlier, a reasoning similar to what has been done above for line $C$ shows that formation of the $B$ and $R$ lines proceeds in exactly the same way, and the resulting expressions for the line intensities are very much the same as Eq.~\ref{eq:integrated_I_Cbcr} above. The reason for this is the intrinsic symmetry of the problem, which is made very clear by  Fig.~\ref{fig:figure2.6}.  We can thus generalize Eq.~\ref{eq:integrated_I_Cbcr} to

\begin{equation}
        I_{B, C, R}^{bcr}(p,z=\infty,\tilde{x}) = I_{B,C,R}^{*}(\tilde{x})e^{-\tau_{B}-\tau_{C}-\tau_{R}}+S_{B}(1-e^{-\tau_{B}})e^{-\tau_{C}-\tau_{R}}+S_{C}(1-e^{-\tau_{C}})e^{-\tau_{R}}+S_{R}(1-e^{-\tau_{R}}). \label{eq:integrated_I_BCR}
\end{equation}
Even though emergent intensities are basically the same for the three lines, their contributions to the emergent profiles are in fact quite different. This is because the frequencies to which each of the resonant regions contributes vary from line to line, as can also be seen in Fig.~\ref{fig:figure2.6}. 

\subsubsection{Line profiles}

In order to show how the interactions affect the line profiles, we display them in two situations: in Fig.~\ref{fig:figure2.7} the lines are computed independently, that is, without taking the interactions discussed above into account\footnote{All line profiles shown in the paper are computed with a numerical resolution of 1km/s.}. Unsurprisingly, the individual profiles display the characteristic features of lines formed in accelerating outflows, with absorption features reaching almost to the rest wavelength of the line and the sharp drop in intensity near zero velocity on the blue line side \citep[see, e.g.,][]{ 1987ApJ...314..726L}. In Fig.~\ref{fig:figure2.8}, we show the emergent profiles when interactions are considered. Line $C$, which has the lowest optical depth of the three lines, is weak in the noninteracting case, but goes into strong emission when interactions are considered. Lines $R$ and $B$ have comparable optical depths, and are strongly enhanced by the interactions. The blends shown in these figures result from co-adding the individual line fluxes.

What can we conclude from this first example of line fluorescence? Clearly, line $C$ is enhanced by photons of the two other lines in an outwardly accelerating outflow. So are $B$ and $R$, and the resulting blend is a wide, triple peaked profile. The blue-displaced absorption characteristic of outflows is essentially that of the $B$ line, somewhat filled up by emission. Given that the interacting line source functions are qualitatively quite similar to what they would be if the lines were independent (compare Figs.~\ref{fig:figure2.4} and \ref{fig:figure2.5}), the gist of line amplification occurs during flux integration. Actually, test computations demonstrated that line profiles obtained with source functions computed simply with the local Sobolev approximation are often qualitatively indistinguishable from those obtained with the interacting source functions used above. Again, this is likely an artifact introduced by the assumption of LTE atomic populations. But it is quite interesting to note that fluorescence caused by line amplification in the observer's frame occurs despite this limiting condition. This is of course a direct consequence of the additional emission in the resonance zones caused by Doppler-shifted neighboring lines in macroscopic velocity fields. 

We next considered a more complex case, one with a different topology for the CRVSs: that of an accretion flow.

\section{Extension to more complex velocity fields\label{sec:extension-to-complex-velocity-fields}}

\begin{figure}
        \centering
        \begin{minipage}{0.48\textwidth}
                \centering
                \includegraphics[width=1.0\textwidth]{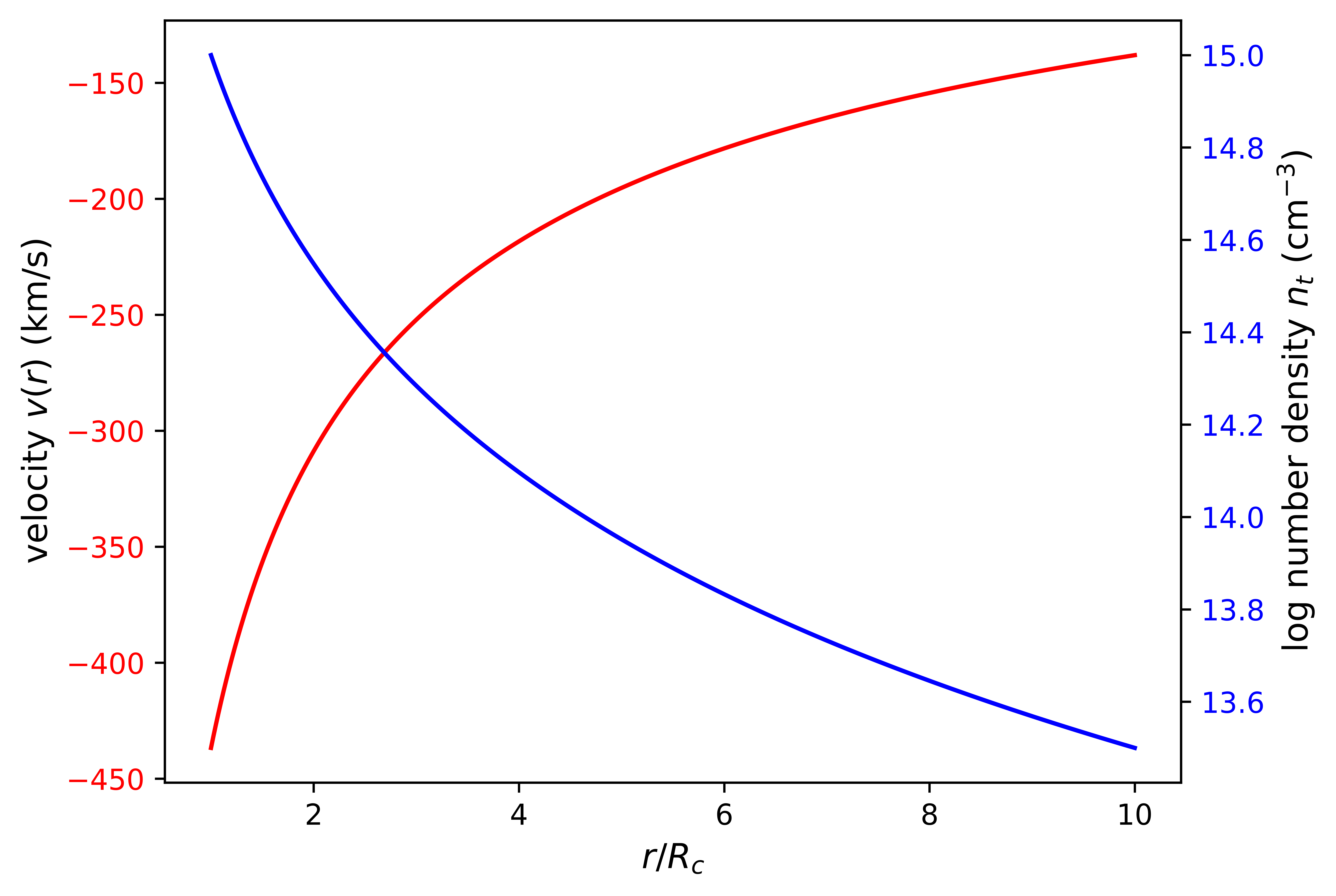} 
                \caption[Short caption]{Velocity and density laws of the accretion flow discussed in Sect. 3. }
                \label{fig:figure4.1}
        \end{minipage}\hfill
        \begin{minipage}{0.48\textwidth}
                \centering
                \includegraphics[width=0.9\textwidth]{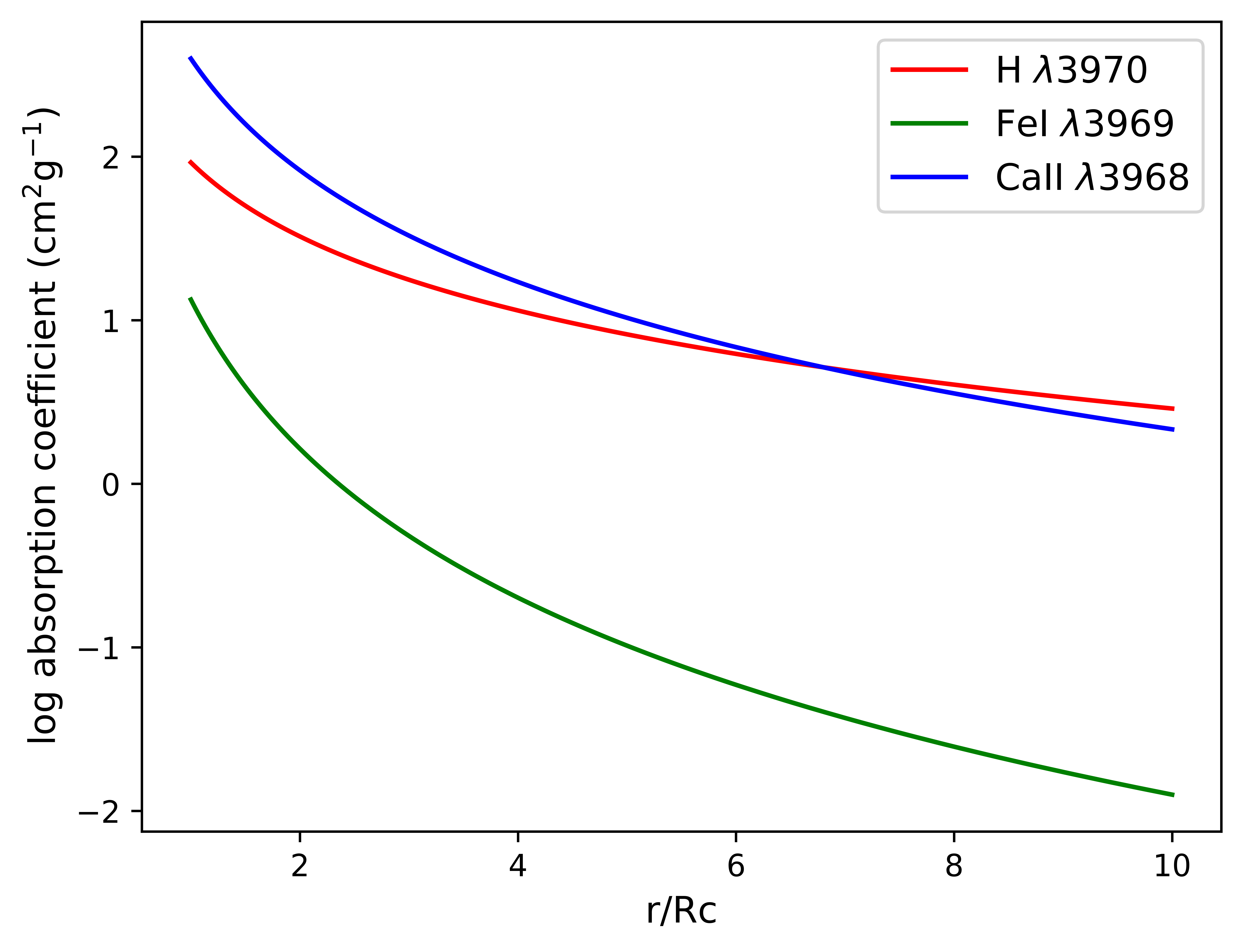} 
                \caption[Short caption]{Line absorption coefficient for the three lines considered.}
                \label{fig:figure4.2}
        \end{minipage}
\end{figure}

\subsection{Accretion flow\label{subsec:accretion-flow}}

Free-falling accretion flows, as well as outwardly decelerating ballistic outflows or rotational flows, all lead to situations where distant parts of the emitting region can interact radiatively even when considering the case of single spectral lines; this is because the constant radial velocity surfaces discussed above are multi-valued along some rays. As we mentioned earlier, a generalization of the Sobolev theory to handle this case was worked out by \citet{1978ApJ...219..654R} for single lines. We recall their main results in the next paragraph before extending the same formalism to interacting lines.

\subsubsection{Sobolev source functions}

We consider an accelerated gravitational accretion flow with velocity law 
\begin{equation}
        v(r) = - V_c \cdot (R_c / r)^{\alpha}, \label{eq:infall}
\end{equation} 
where $V_c$ is the free-fall velocity at $R_c$. The computation parameters are summarized in Table~\ref{tab:table2}. Since the maximum envelope gas density is 20 times lower than in the case discussed in the previous section, the line absorption coefficients of Fig.~\ref{fig:figure4.2} are correspondingly smaller, but sufficient for producing sizable line emission in this velocity field, as seen below.

\begin{table}[h!] \caption{Computation parameters for the accretion flow of Sect.~\ref{sec:extension-to-complex-velocity-fields}.} \label{tab:table2}
        \begin{tabular}{lllllllll}
                \toprule
                R$_c$& R$_{env}$ & M$_* $ &     \teff & \tenv &  $n_t(R_c)$ & $V_{c}$ &     $\alpha$ \\
                \midrule
                2 R$_\sun$ & 10 R$_c$ & 1 M$_\sun$ & 4.5 $\cdot 10^3$K & $ 6 \cdot 10^{3}$ K & $1 \cdot 10^{15}$ cm$^{-3}$ & 436.76 km/s  & 0.5 \\
                \bottomrule
        \end{tabular}
\end{table}

Since the envelope velocity is varying monotonically, the photons of the reddest considered line $R$ are seen redshifted by atoms emitting bluer line photons, so that the source function of line $R$ remains unaffected by the other transitions considered. We therefore start our discussion with transition $R$, to which the  \citet{1978ApJ...219..654R} equations directly apply.

\paragraph{$R$ transition.} 

\begin{figure*}[tp]
        \centering
        \includegraphics[width=1.0\linewidth]{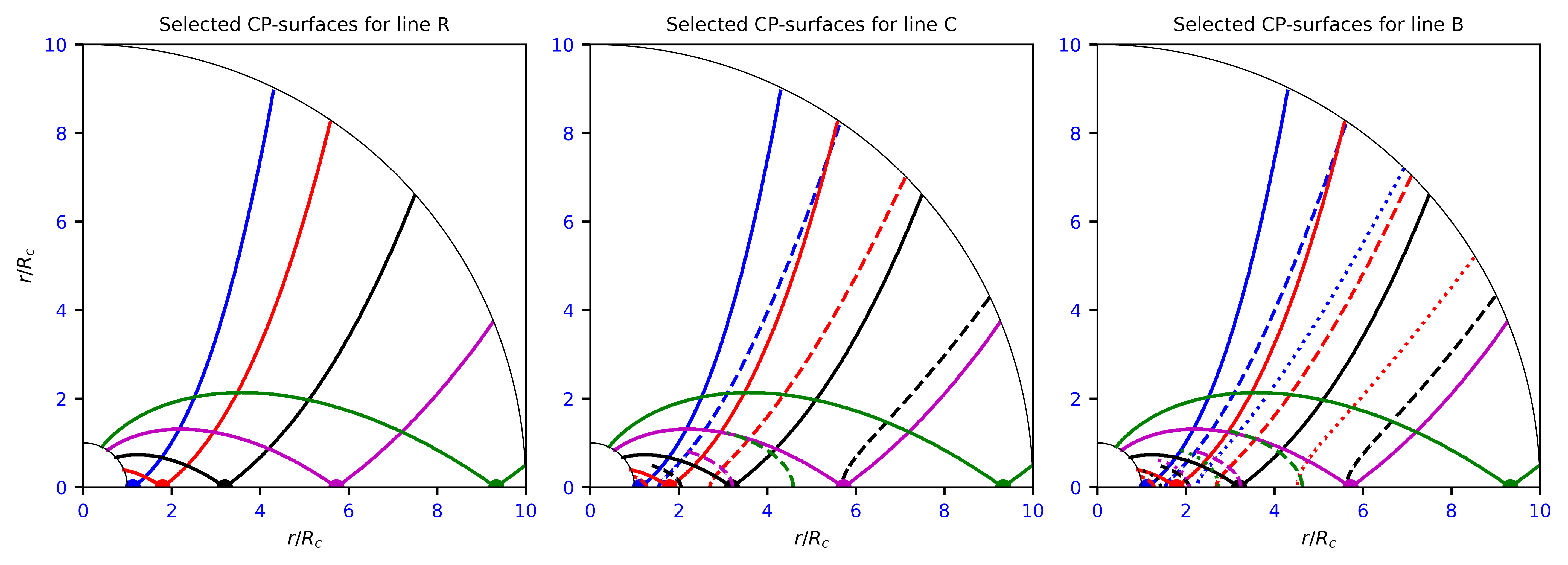}
        \caption[Short caption]{Selected CP surfaces for the spherically symmetric accretion flow, with properties given in the text. Each surface is associated with the radial point of the same color.}
        \label{fig:figure4.3}
\end{figure*}

The generalized Sobolev source function for the case of an equivalent two-level atom in an inwardly accelerating radial flow is given by 
\begin{equation}
        S_{R}(r)=\frac{(1-\varepsilon_{R})\bar{\beta}_{R}^{c}I_{R}^{c}+\varepsilon_{R}B_{R}+(1-\varepsilon_{R})S_{R}^{nl}}{\varepsilon_{R}+(1-\varepsilon_{R})\beta_{R}}, \label{eq:S_gen}
\end{equation}
with
\begin{equation}
        S^{nl}_{R}(r) = \frac{1}{2}\int_{-1}^{+1}\frac{1-e^{-\tau_{R}}}{\tau_{R}}(1-e^{-\tau(r^{*}_R)})S(r^{*}_R)d\mu,
\end{equation}
where $\tau(r^{*}_R)$ and $S(r^{*}_{R})$ are evaluated at the CP surface $r^{*}_{R}$ associated with $O(r)$. The third term of Eq. \ref{eq:S_gen} represents the main nonlocal contribution to the source function that arises in these flows. A second nonlocal contribution appears in the term $\bar{\beta}_{R}^{c}I_{R}^{c},$ which represents the contribution of the stellar continuum flux at line wavelength, attenuated by the intervening CP layers between core and $O(r)$:
\begin{equation}
        \bar{\beta}_{R}^{c}I_{R}^{c}=\frac{I_{R}^{c}}{2}\intop_{\mu_{c}}^{+1}\frac{1-e^{-\tau_{R}}}{\tau_{R}}e^{-\tau(r^{*}_R)}d\mu.
\end{equation}
The left panel of Fig.~\ref{fig:figure4.3} shows the CP surfaces associated with line $R$ (in this case the \ce{H_$\epsilon$} line), which can be compared to the CP surfaces found by  \citet{1978ApJ...219..654R}. These are self-CP surfaces in the sense that transition $R$ interacts radiatively with distant photons at the same rest wavelength. 

\begin{figure}
        \centering
        \begin{minipage}{0.48\textwidth}
                \centering
                \includegraphics[width=0.95\textwidth]{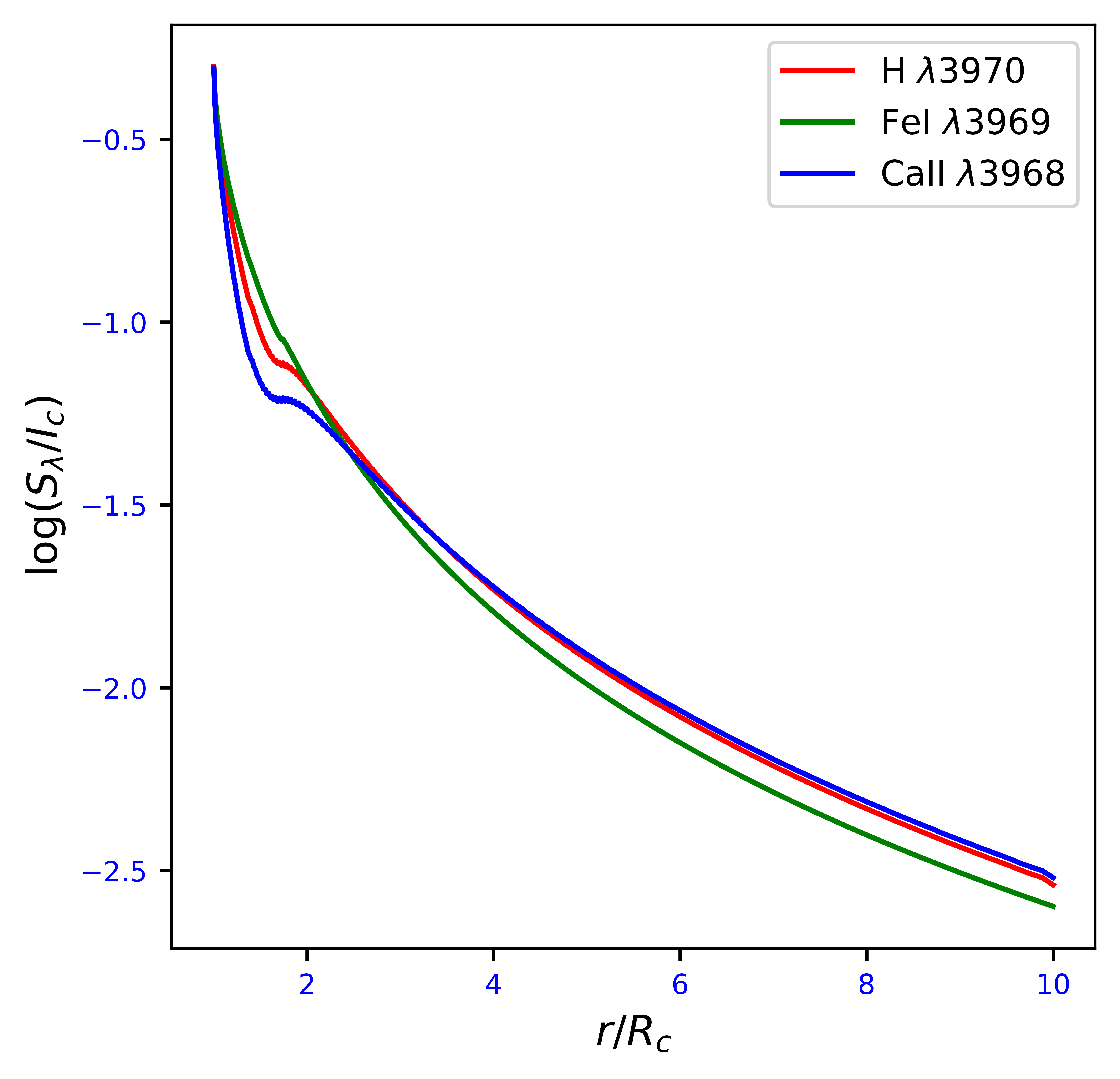} % first figure itself
                \caption[Short caption]{Source functions for the three lines when they are computed independently.}
                \label{fig:figure4.4}
        \end{minipage}\hfill
        \begin{minipage}{0.48\textwidth}
                \centering
                \includegraphics[width=0.95\textwidth]{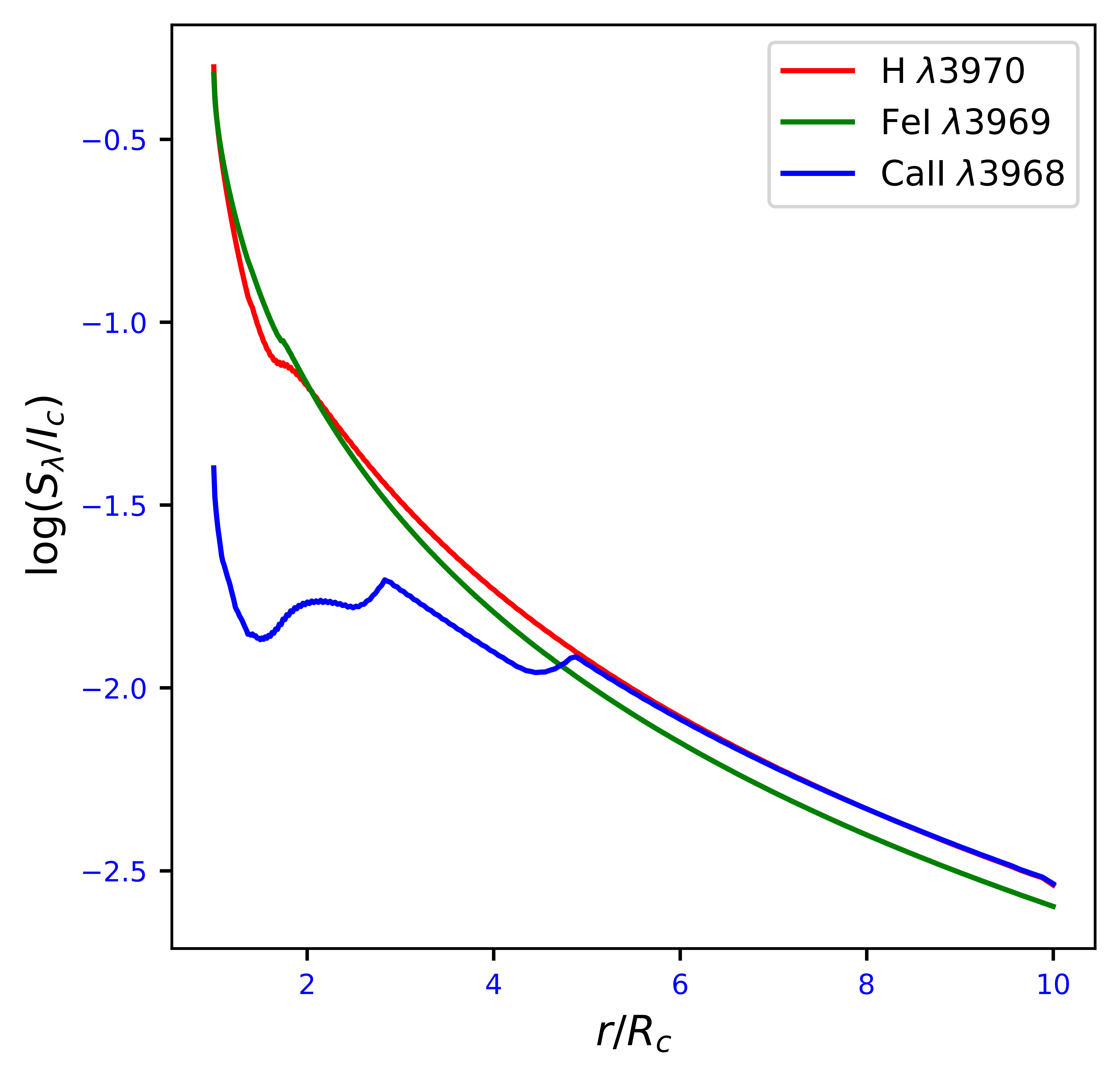} % second figure itself
                \caption[Short caption]{Source functions for the three lines when radiative interactions are taken into account.}
                \label{fig:figure4.5}
        \end{minipage}
\end{figure}

The analogy of these equations with the case of interacting lines $B$ and $C$ (Eqs.~\ref{eq:SC}~to~\ref{eq:BetaCc}) of the previous section is obvious. In the previous case of transition~$C$, the integration ran over a resonance surface corresponding to the locus of $B$ photons that were Doppler-shifted by the right amount, providing an additional source term that represents a net gain of photons for transition~$C$. For a single line formed in an accretion flow, on the other hand, the local source term for $R$ is supplemented by nonlocal photons of the same transition $R$, originating from the CP surface associated with the considered emission point in the envelope.

\paragraph{$C$ and $B$ transitions.} 
The expressions for $S_{C}(r)$ and $S_{B}(r)$ are similar to Eq.~\ref{eq:S_gen} but with the following changes in the nonlocal parts of the source function, reflecting the fact that the self-CP surface $r^{*}$ of the line  is not the only region that must be taken into account in the nonlocal integral. Indeed, atoms emitting line $C$  will be able to absorb some blue-displaced line-$R$ photons, while atoms emitting transition $B$ will interact with blue-displaced photons from both $C$ and $R$. Figure~\ref{fig:figure4.3}, in its center and right panels, show the set of CP surfaces that are contributing to the nonlocal parts of the respective source functions. The graphical convention is the following: solid lines indicate line self-interactions, dashed lines show the CP surfaces corresponding to the nearest transition, and dotted lines indicate the CP surfaces corresponding to the third, farthest transition. 

As previously, the condition of interaction for the $C$ line is given by
\begin{equation}
        v_{rad}(r_{R})-v_{rad}(r_{C})=\Delta\lambda_{RC}/c,
\end{equation}   
which defines a CP surface denoted by $r^{'}_{R}$ in the following expressions. With $\tau(r^{'}_{R}) = \tau^{'}_{R}$ and $\tau(r^{*}_{C}) = \tau^{*}_{C}$, we have
\begin{equation}
        S_{C}(r)=\frac{(1-\varepsilon_{C})\bar{\beta}_{C}^{c}I_{C}^{c}+\varepsilon_{C}B_{C}+(1-\varepsilon_{C})S_{C}^{nl}}{\varepsilon_{C}+(1-\varepsilon_{C})\beta_{C}}, \label{eq:S_C_gen}
\end{equation}
with
\begin{equation}
        S^{nl}_{C}(r) = \frac{1}{2}\int_{-1}^{+1}\frac{1-e^{-\tau_{C}}}{\tau_{C}}\left [(1-e^{-\tau^{*}_{C}})S(r^{*}_C)+(1-e^{-\tau^{'}_{R}})S(r^{'}_R) e^{-\tau^{*}_{C}}\right ] d\mu
\end{equation}
and
\begin{equation}
        \bar{\beta}_{C}^{c}=\frac{1}{2}\intop_{\mu_{c}}^{+1}\frac{1-e^{-\tau_{C}}}{\tau_{C}} e^{-(\tau^{*}_{C} {+\tau^{'}_{R}})}d\mu.
\end{equation}

\noindent For line~$B$, two resonance conditions on the radial velocity lead to line interactions, namely
\begin{equation} \left\{
        \begin{array}{l}                
                v_{rad}(r_{B})-v_{rad}(r_{C})=\Delta\lambda_{BC}/c \\
                v_{rad}(r_{B})-v_{rad}(r_{R})=\Delta\lambda_{BR}/c \label{eq:CP_eq4}            
        \end{array} \right. \
.\end{equation}

\begin{figure*}[tp]
%\centering
\includegraphics[width=1.0\linewidth]{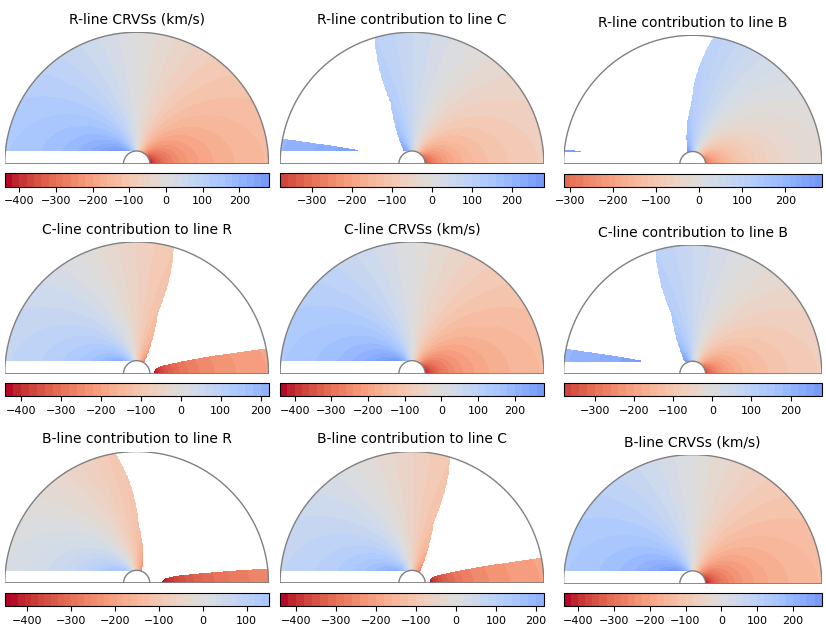}
\caption{Regions contributing to the formation of the considered line profiles in the accretion flow of Sect.~4. See the caption of Fig.~\ref{fig:figure2.6} for details.}
\label{fig:figure4.6}
\end{figure*}

\begin{figure}[h]
        \centering
        \begin{minipage}{0.48\textwidth}
                \centering              \includegraphics[width=0.95\textwidth]{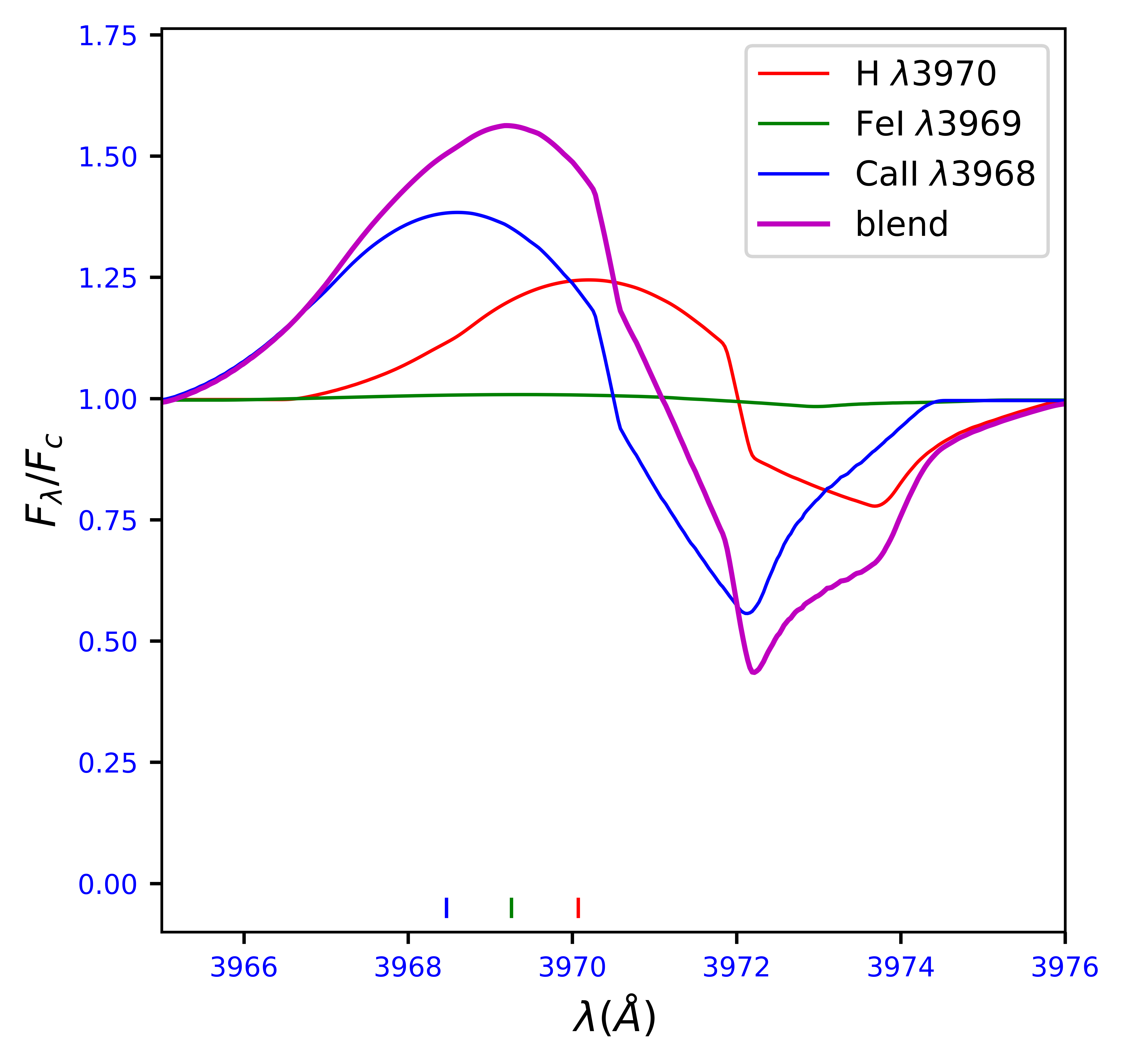} % first figure itself
                \caption[Short caption]{Line profiles of the three lines considered and their blend for the accretion flow discussed here when the lines are computed independently.}
                \label{fig:figure4.7}
        \end{minipage}\hfill
        \begin{minipage}{0.48\textwidth}
                \centering
                \includegraphics[width=0.95\textwidth]{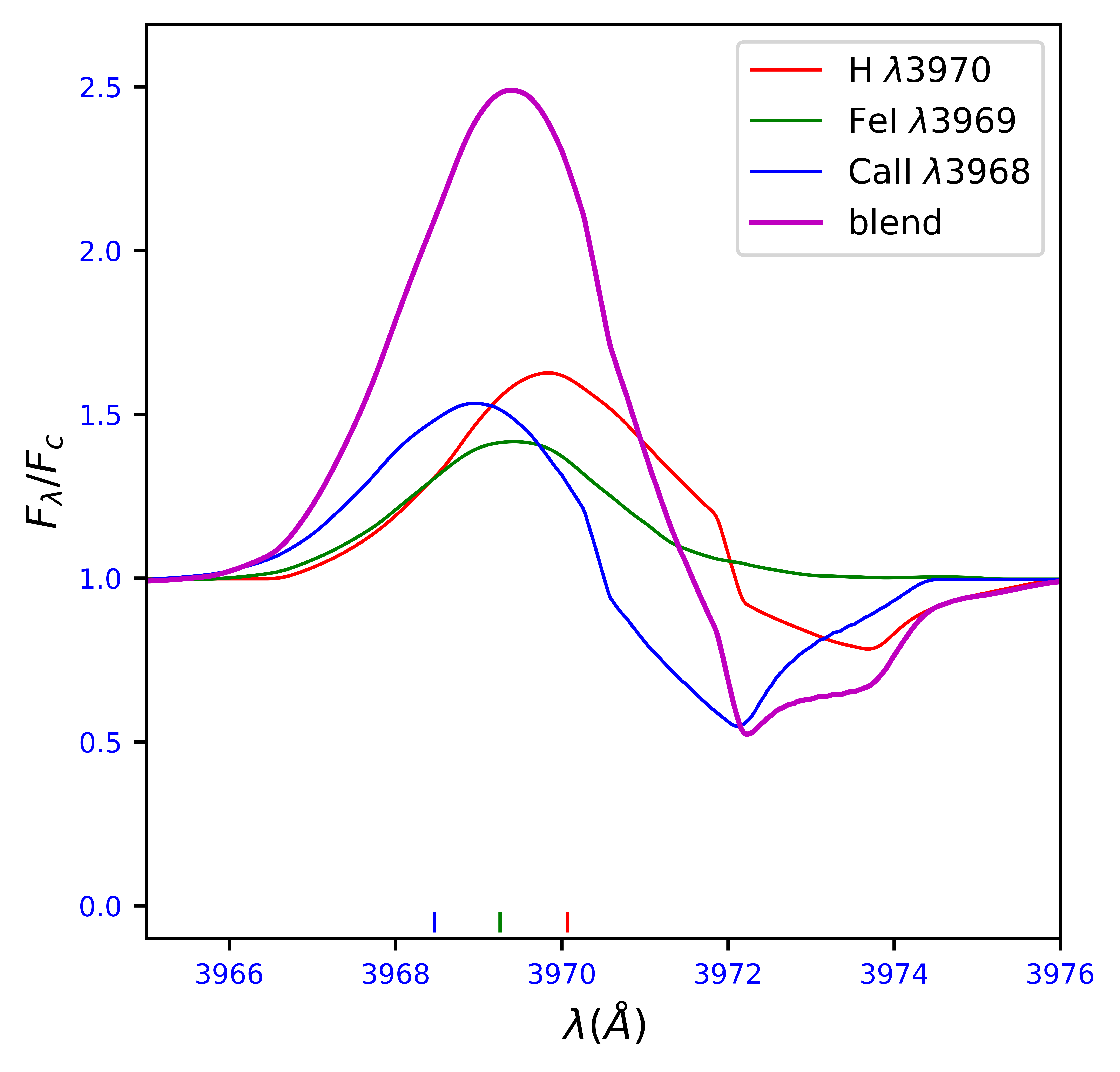} % second figure itself
                \caption[Short caption]{Line profiles of the three lines considered and their blend for the accretion flow discussed here when radiative interactions between lines are taken into account.}
                \label{fig:figure4.8}
        \end{minipage}
\end{figure}

We must therefore locate the resonance regions corresponding to these conditions, the loci of which are noted $r^{''}_{C}$ and $r^{'''}_{R}$ in the following equations. In a direct extension of Eq.~\ref{eq:SR}, we then have
\begin{equation}
        S_{B}(r)=\frac{(1-\varepsilon_{B})\bar{\beta}_{B}^{c}I_{B}^{c}+\varepsilon_{B}B_{B}+(1-\varepsilon_{B})S_{B}^{nl}}{\varepsilon_{B}+(1-\varepsilon_{B})\beta_{B}}, \label{eq:S_B_gen_1}
\end{equation}
with
\begin{equation}
        S^{nl}_{B}(r) = \frac{1}{2}\int_{-1}^{+1}\frac{1-e^{-\tau_{B}}}{\tau_{B}}
        \left [(1-e^{-\tau^{*}_{B}})S(r^{*}_B)+(1-e^{-\tau^{''}_{C}})S(r^{''}_C) e^{-\tau^{*}_{B}}+(1-e^{-\tau^{'''}_{R}})S(r^{'''}_R) e^{-(\tau^{*}_{B}+\tau^{''}_{C})}\right ] d\mu \label{eq:S_B_gen_2}
\end{equation}
and
\begin{equation}
        \bar{\beta}_{B}^{c}=\frac{1}{2}\intop_{\mu_{c}}^{+1}\frac{1-e^{-\tau_{B}}}{\tau_{B}} e^{-(\tau^{*}_{B}{+\tau^{''}_{C}}+{\tau^{'''}_{R})}}d\mu. \label{eq:S_B_gen_3}
\end{equation}

The interaction matrix for this case  must include the self-interaction of each line, so that it becomes, with line ordering $R, C, B$,
\begin{equation}
        I = \left(\begin{array}{ccc}
                1 & 0 & 0 \\ 1 & 1 & 0 \\ 1 & 1 & 1 
        \end{array}\right).
\end{equation}
Figure~\ref{fig:figure4.4} displays the computed source functions for the three lines in consideration when they are treated as independent, that is, when Eq.~\ref{eq:SB} is used for computing all of them. Figure~\ref{fig:figure4.5} shows the source functions computed when taking the radiative interactions into consideration, that is, when using Eq.~\ref{eq:S_C_gen} for the $C$ line and Eq.~\ref{eq:S_B_gen_1} for the $B$ line. The changes caused by core photon absorption at the CP surfaces  are small for the optically thin \ce{Fe I} line but lead to a strong depression of the optically thick \ce{Ca II} line source function below $R \sim 4$. We recall that the \ce{H_\epsilon} line source function is the same in both cases. 

\subsubsection{Flux integration}

In an accelerating infall or decelerating outflow, the unique correspondence between line frequencies and positions on the line of sight found in accelerating outflows is lost, as can be seen in the top left panel of Fig.~\ref{fig:figure4.6}, which displays the envelope's gas radial velocities seen by an outside observer located at $z = \infty$. Depending on the impact parameter and envelope extent, the CRVSs are either open or closed, so that there are one or two locations on the line of sight that have the same frequency displacement. As explained in \citet{1984ApJ...285..269B}, the flux integration method that we use handles this situation without modification. It finds all line frequency displacements present on a given impact parameter and orders them for the line intensity computation. 

The line flux of $R$ is made up of several contributions shown in the first column of Fig.~\ref{fig:figure4.6}. They are first the line-$R$ photons originating in the CRVSs shown in the top panel, then the displaced $C$-line photons originating in deeper parts of the envelope (middle panel), and also  the displaced $B$-line photons originating even deeper in the envelope (bottom panel). The final contribution to the red side of line $R$ comes from $C$ and $B$ photons emitted by fast-moving atoms located in front of the star. 
As illustrated by the second column of Fig.~\ref{fig:figure4.6}, line~$C$ gets a contribution from~$R$ photons that are emitted in the outside part of the envelope (as seen by the observer) where atoms are moving slower by an amount $\delta v$, so the flux is enhanced by emission from the front part of the envelope. Line~$C$ gets an additional contribution from~$B$ photons that are coming from the deeper parts of the envelope and displaced by the right amount. We again note the additional contribution to the $C$-line red wing from  $B$ photons emitted by fast-moving atoms located in front of the star. Similar considerations hold for line $B$, and the various origins of its flux are shown in the right column of Fig.~\ref{fig:figure4.6}. 

Figures~\ref{fig:figure4.7} and \ref{fig:figure4.8} display the computed line profiles for two situations. In Fig.~\ref{fig:figure4.7} the lines are computed as if they were single lines, without taking all interactions discussed above into account, while in Fig.~\ref{fig:figure4.8} we show the profiles that result when interactions are considered. Profiles of Fig. ~\ref{fig:figure4.7} are typical accretion flow profiles with intermediate optical depths. Line $C$ has the lowest optical depth of the three lines; it is indistinguishable from the continuum when the lines are computed independently but goes in moderate emission when interactions are taken into account. Line $R$ has a somewhat lower optical depth than $B$ but is enhanced by contributions of photons from the other transitions. The flux of these last two lines are similar, so the blend is in emission, with a single peak approximately centered on the strongest line $R$; it displays a moderately deep but wide absorption component that results from a combination of the three individual absorption features.

\subsection{Non-monotonic velocity fields\label{subsec:non-monotonic-flow}}

\begin{figure}[t]
        \centering
        \begin{minipage}{0.48\textwidth}
                \centering
                \includegraphics[width=0.95\textwidth]{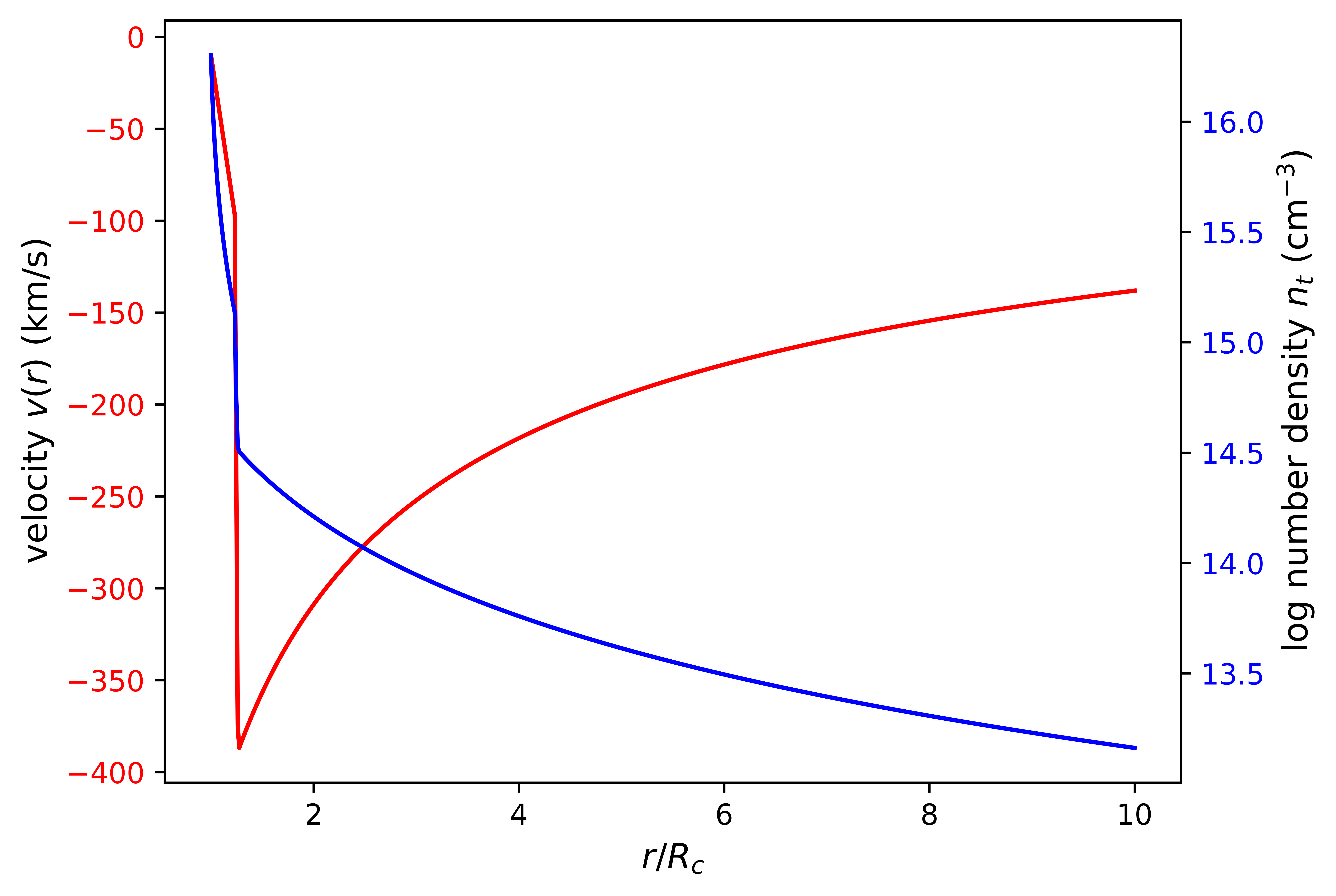} 
                \caption[Short caption]{Velocity and density laws of the non-monotonic accretion flow discussed in this section.}
                \label{fig:figure4.15}
        \end{minipage}\hfill
        \begin{minipage}{0.48\textwidth}
                \centering
                \includegraphics[width=0.95\textwidth]{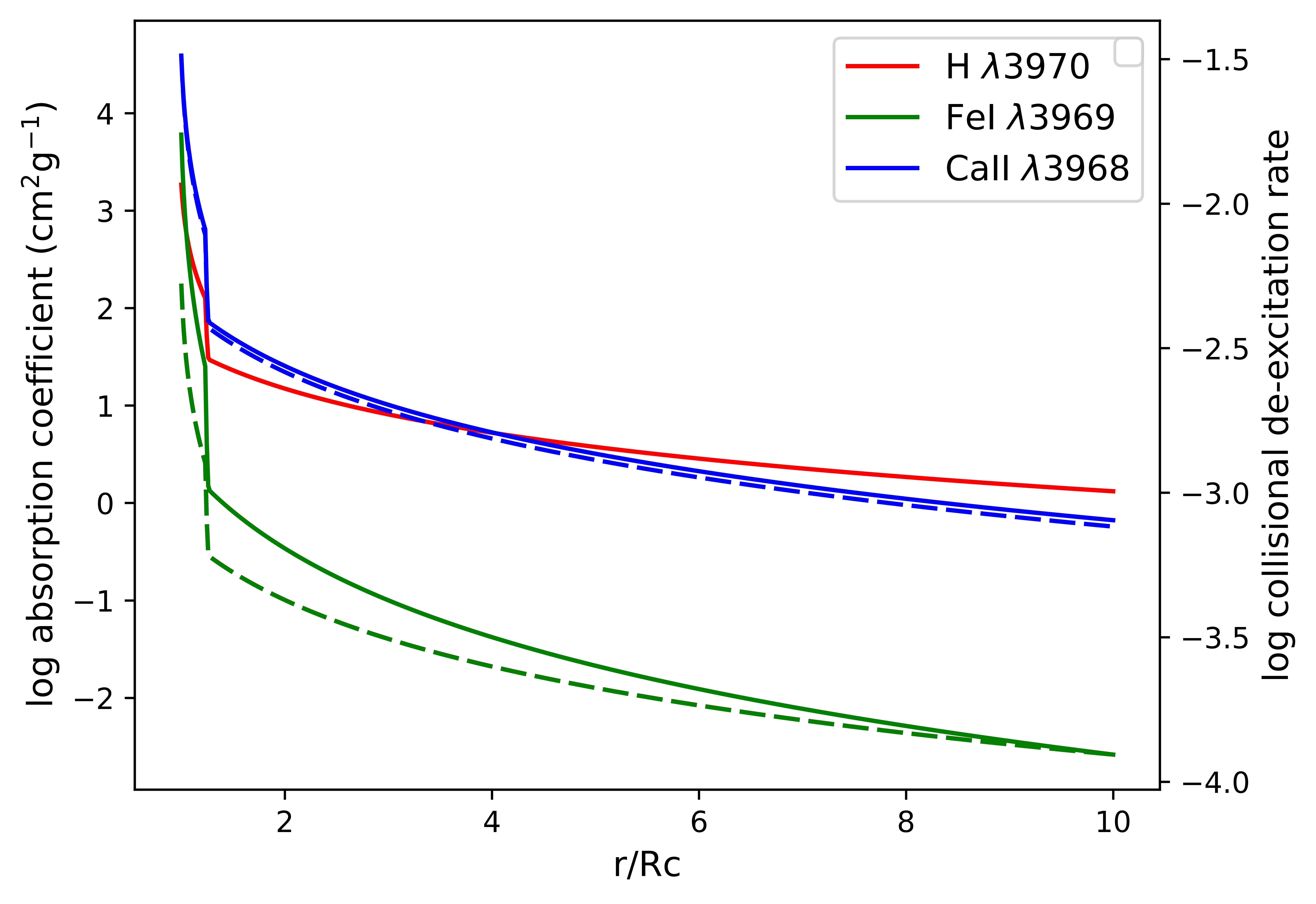} 
                \caption[Short caption]{Line absorption coefficient and rates of collisional de-excitation for the three lines considered.}
                \label{fig:figure4.16}
        \end{minipage}
\end{figure}

The method developed here readily extends to non-monotonic velocities, as found, for example, in an accreting envelope with an accretion shock close to the stellar surface. To illustrate this,
we considered a qualitative TTS accretion shock toy model, the velocity and density fields of which are displayed in Fig.~\ref{fig:figure4.15}. The gas free-fall velocity increases toward the star with the usual $v(r) \propto r^{-1/2}$ radial dependence until it reaches the  shock at $R_s$, assumed to be located at 1.2 $R_c$. The shock is assumed to be isothermal, so the gas velocity decreases by a factor of 4 at $R_s$. This jump is followed by an (assumed) linear deceleration and reaches the stellar radius $R_c$ at 10~km/s. The density follows from the continuity equation, and the gas density at the shock is $n_t(R_s) = 5 \cdot 10^{15}$ cm$^{-3}$. We did not consider the complex physics involved in the shock, but simply solved the formation of our three lines with these values of velocity, temperature, and density fields.

\begin{figure*}
        \centering
        \includegraphics[width=1.0\linewidth]{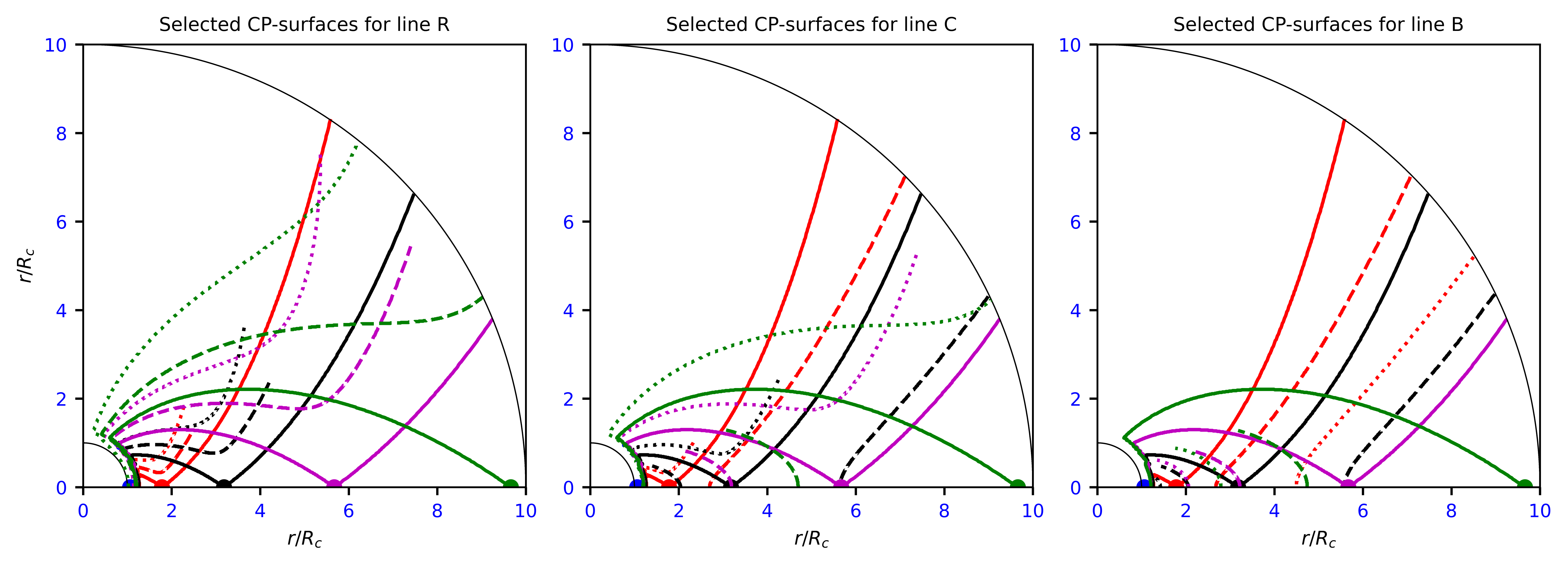}
        \caption[Short caption]{Selected CP surfaces for the spherically symmetric non-monotonic accretion flow, with properties given in the text. Each surface is associated with the radial point of the same color.}
        \label{fig:figure4.17}
\end{figure*}

For more realistic source functions, we introduced collisional de-excitation rates in an approximate way by using formulas from Allen's Astrophysical Quantities \citep{2000asqu.book.....C}, which actually go back to \citet{1962ApJ...136..906V}. Clearly, more modern data should be used for a detailed comparison with observations, but at this preliminary stage a crude estimate of the collision effects is sufficient. The collisional de-excitation rates (dashed lines) and the absorption coefficients (solid lines) of the three lines are shown in Fig.~\ref{fig:figure4.16}. The collisional de-excitation rate of the calcium ion line is an order of magnitude higher than the rates of neutral hydrogen and iron lines, which are indistinguishable in this graph.  The three lines are moderately optically thick in the post-shock region, where much of the fluorescent amplification occurs.

\begin{figure}
        \centering
        \begin{minipage}{0.48\textwidth}
                \centering              \includegraphics[width=0.95\textwidth]{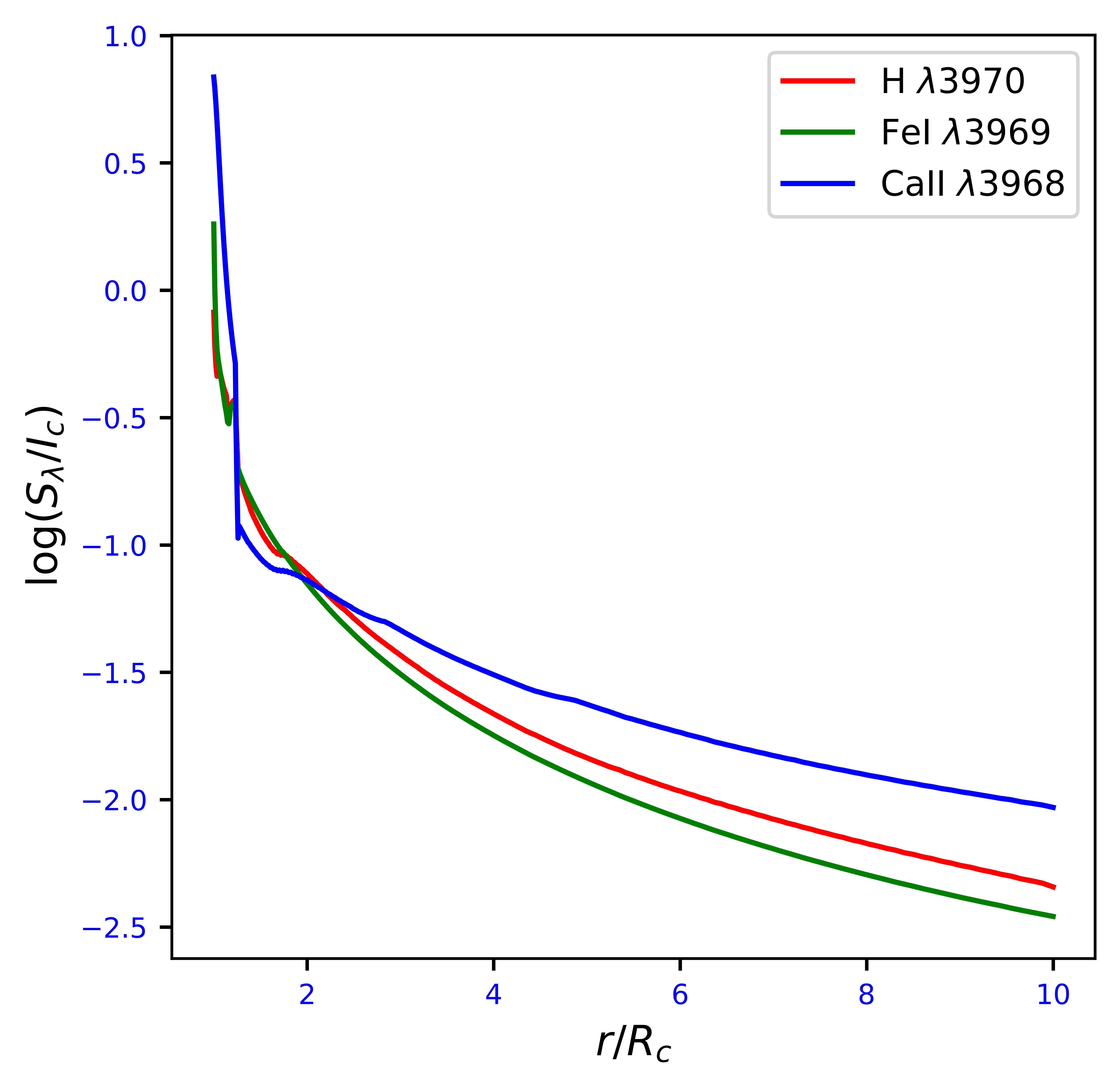} % first figure itself
                \caption[Short caption]{Source functions of the three lines considered and their blend for the non-monotonic accretion flow discussed here.}
                \label{fig:figure4.19}
        \end{minipage}\hfill
        \begin{minipage}{0.48\textwidth}
                \centering
                \includegraphics[width=0.95\textwidth]{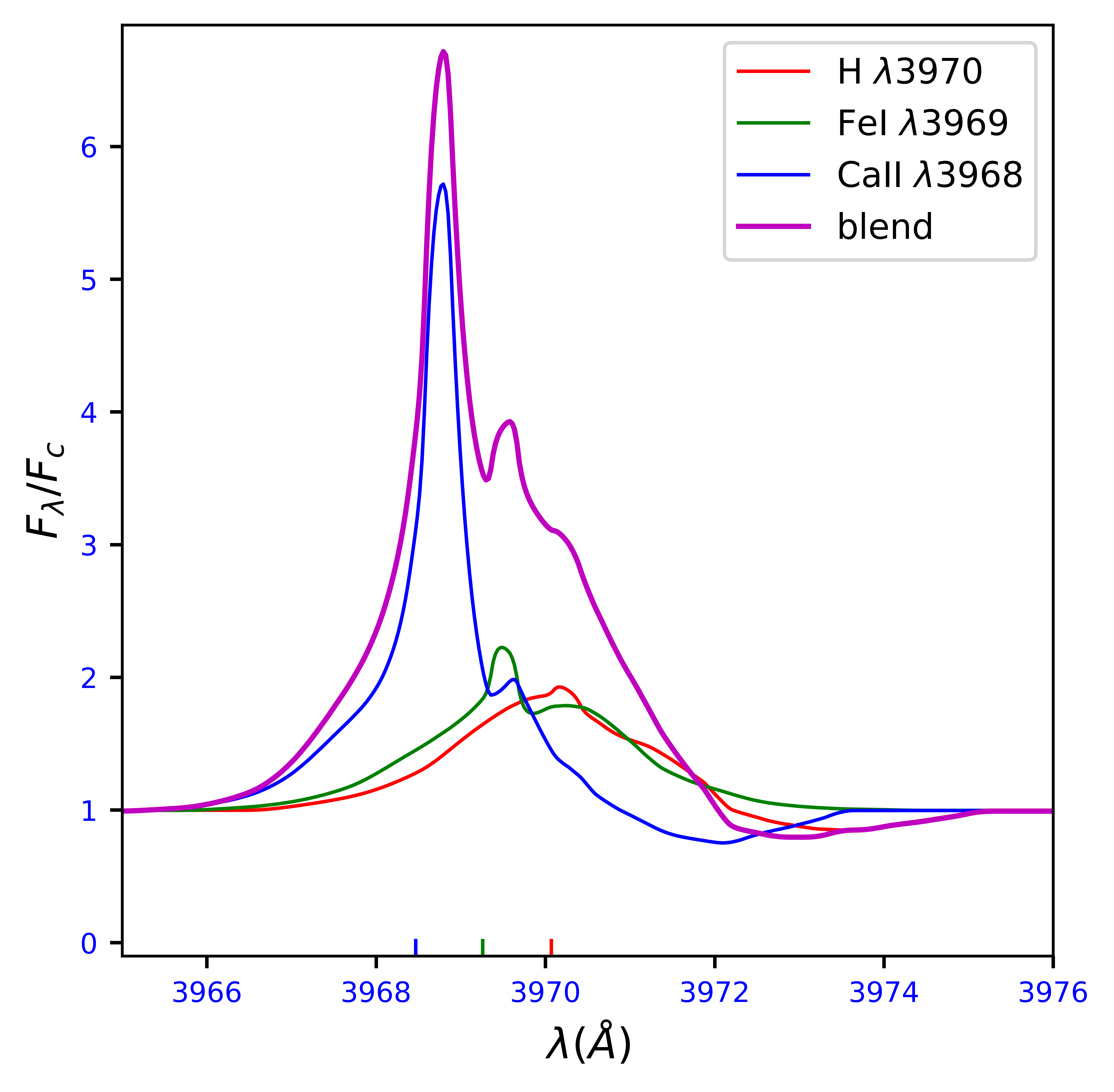} % second figure itself
                \caption[Short caption]{Line profiles of the three lines considered and their blend for the non-monotonic accretion flow discussed here.}
                \label{fig:figure4.20}
        \end{minipage}
\end{figure}

The CP surfaces for the three lines considered are shown in Fig.~\ref{fig:figure4.17}. As in previous panels showing CPs, the solid lines display the self-CP surfaces of each line, while the dashed lines show the interaction surfaces with the first neighboring line and the dotted lines display the interaction surfaces with the second neighboring line. We see that, unlike the monotonic cases, all lines are, as expected, involved in the interactions, and the CP surfaces are a mix of those found in the previous cases of monotonic flows. They introduce discontinuities in the source functions, particularly in the shock region, as can be seen in Fig.~\ref{fig:figure4.19}.

The resulting line profiles are displayed in Fig.~\ref{fig:figure4.20}. They do not resemble those found for the accretion flow of the previous section, partly because most of the flux originates in the post-shock deceleration region close to the star, and partly because of the collision-term effect. Despite the fact that the optical depths of the \ce{H} and \ce{CaII} lines are very similar, the emergent profiles are quite different, with \ce{CaII}~H showing a steep central peak caused by the collision term that enhances its source function in the innermost envelope region (see Fig.~\ref{fig:figure4.19}).
We note that the three lines participate in the fluorescence process, with the iron line getting contributions from both neighboring lines and thus being amplified most.

\section{Discussion and conclusion}\label{sec:Discussion}

\begin{figure}[]
        \centering
        \includegraphics[width=1.0\linewidth]{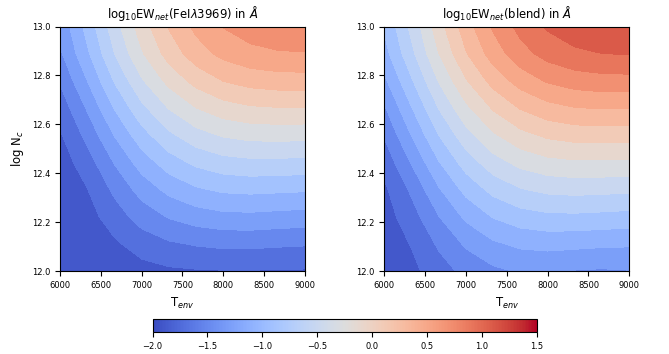}
        \caption[Short caption]{Parameter space study for the net EWs of \ce{FeI}$\lambda3969$ line and its blend with H and H$_\epsilon$, in the case of an accretion flow with $\epsilon \neq 0$.}
        \label{fig:figure4.22}
\end{figure}

We have investigated a rather small parameter space in the previous sections of this paper, mainly because we required that \ce{CaII} H and K have comparable strengths, as is observed in many TTSs. As discussed in Sect.~\ref{sec:accelerated-outflow}, this requires -- when LTE and isothermal envelope are assumed -- that gas temperatures be restricted to a limited range around 6000K, which in turn implies, when ignoring collisional excitation, high gas densities at the envelope's bottom and resulting mass flow rates that appear unrealistically high for TTSs.

\begin{table}[h!] \caption{Computation parameters for the parameter space study discussed here} \label{tab:table3}
        \begin{tabular}{lllllllll}
                \toprule
                R$_c$& R$_{env}$ & M$_* $ &     \teff & \tenv &  $n_t(R_c)$ & $V_{c}$ &     $\alpha$ \\
                \midrule
                2 R$_\sun$ & 10 R$_c$ & 0.5 M$_\sun$ & 3.0 $\cdot 10^3$K &  $10^{12} - 10^{14}$cm$^{-3}$ K & $6 \cdot 10^3 - 9 \cdot 10^3$K & 436.76 km/s  & 0.5 \\
                \bottomrule
        \end{tabular}
\end{table}

 \begin{figure}
        \centering
        \begin{minipage}{0.48\textwidth}
                \centering              \includegraphics[width=0.95\textwidth]{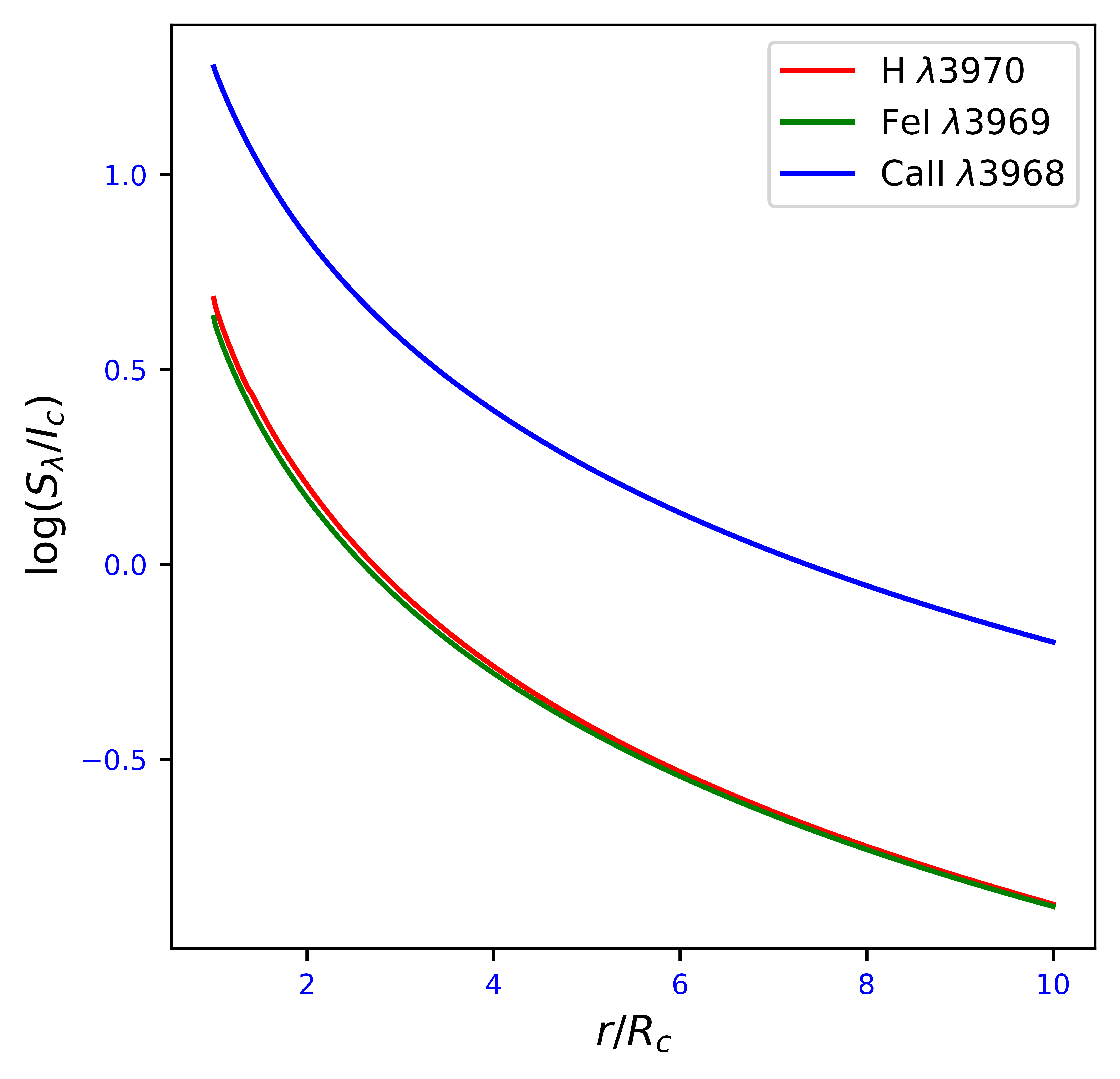} % first figure itself
                \caption[Short caption]{Source functions of the three lines considered for an accretion flow with $n_t(R_c) = 6 \cdot 10^{12}$cm$^{-3}$ and T$_{env} = 8.5 \cdot 10^3$K when collisional de-excitation rates are included.}
                \label{fig:figure4.23}
        \end{minipage}\hfill
        \begin{minipage}{0.48\textwidth}
                \centering
                \includegraphics[width=0.95\textwidth]{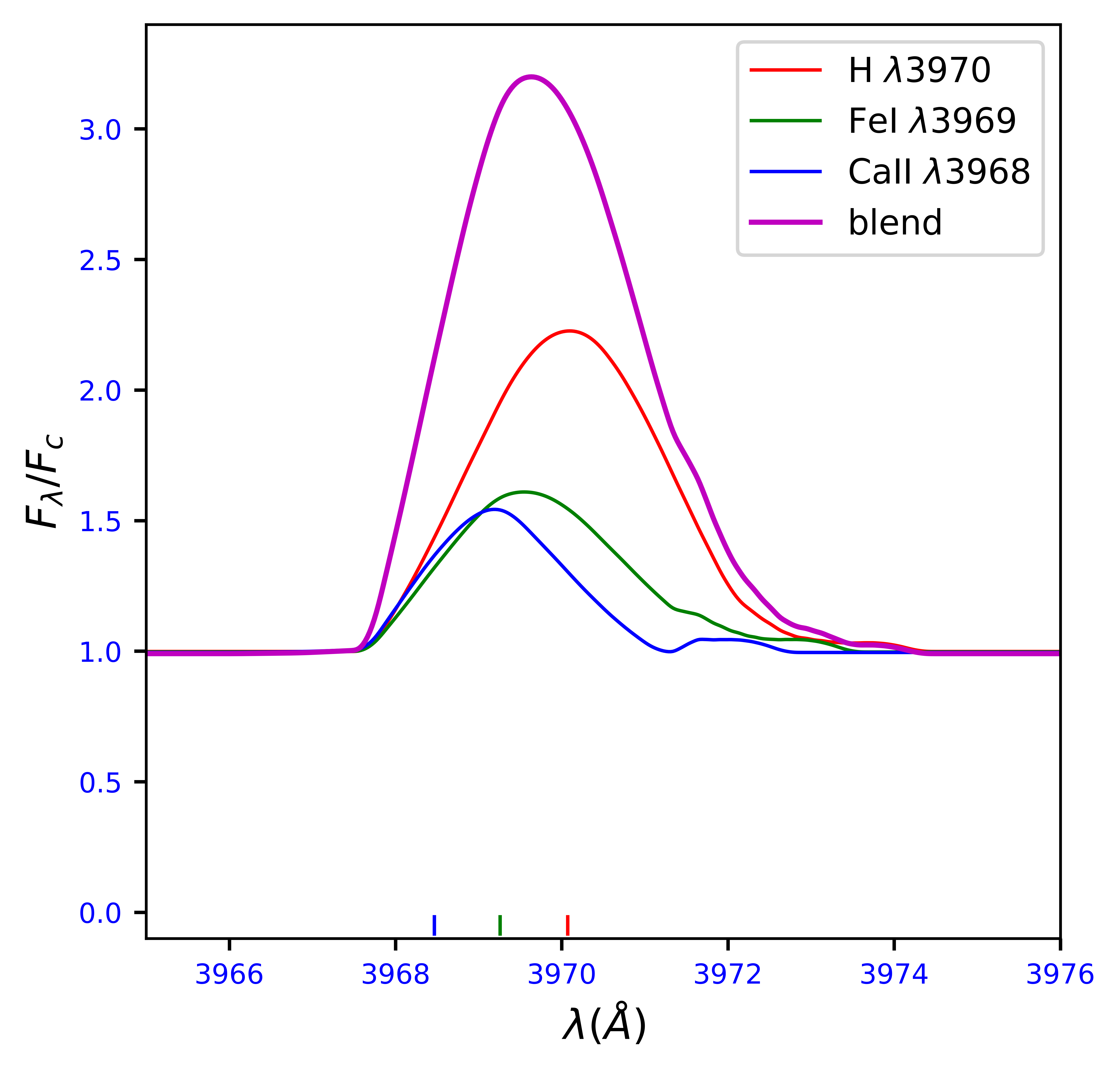} % second figure itself
                \caption[Short caption]{Line profiles of the three lines considered for an accretion flow with the properties given in the previous figure caption.} %Fig.~\ref{fig:figure4.23}.
                \label{fig:figure4.24}
        \end{minipage}
\end{figure}

However, fluorescence is not limited to this narrow range of parameters, but can also take place at higher temperatures and at lower densities when we include the collisional de-excitation terms in the source functions.  This is because the collision term enhances the source function, leading to sizable flux for low line optical depths. Figure~\ref{fig:figure4.22} shows the result of a parameter space exploration that includes collisional de-excitation rates in the source function computation. There, we display the net equivalent widths (EWs) of both  \ce{FeI}$\lambda$3969 and the blend of the three lines, as computed in a more extended parameter space investigation. We vary $n_t(R_c)$ and \tenv\ in the ranges indicated in Table~\ref{tab:table3}, which also gives the other computation parameters. The collisional terms are included in the approximation discussed earlier. The net EW\footnote{By convention, emission line components have positive EWs here. } of a line is defined as its computed EW when the computation includes interactions with the other two lines, from which we subtracted a baseline EW computed under the assumption that the three lines are independent. The net EW is thus a measure of the fluorescent flux. 

In Fig.~\ref{fig:figure4.22} we plot the logarithm of the net EWs, which ranges from -2 to +1.5 in the parameter space investigated. We find that a significant \ce{FeI}$\lambda$3969 fluorescent flux (EW$_{net} \gtrapprox 1\AA$) is produced over a wide range of temperatures for $n_t(R_c) \gtrapprox 4 \cdot 10^{12}$cm$^{-3}$. The interacting lines at these relatively low densities are all optically thin, with pure, almost featureless emission profiles driven by the collision terms; the sources and profiles of a representative example are shown in Figs.~\ref{fig:figure4.23} and \ref{fig:figure4.24}.  The minimum density above corresponds to a spherical mass flow rate of $\approx 10 ^{-6}$M$_\sun$/yr. This mass accretion rate value remains quite high for TTSs in general, but it may represent a plausible value for the most active members of the class  \citep{1995ApJ...452..736H}. One should also recall that TTSs are highly variable, which indicates that the accretion process is not steady \citep [cf.] [] {2007prpl.conf..479B}. Fluorescence might therefore occur in periods of increased accretion but disappear during more quiescent phases, as observed recently in VW~Cha (A. Armeni, priv. comm.).  

As an absolute calibration of TTS mass accretion rates is notoriously difficult \citep [see the discussion in][] {1998ApJ...492..323G}, it seems worthy to improve the atomic model and try computing the precise threshold mass accretion rate for fluorescence to set in. This will provide a useful new  comparison tool for other mass flow rate determination methods.
\smallskip

\noindent With this work, we are now in a position to compute the formation of fluorescent lines in spherically moving media. However, as already mentioned, we are still far from being able to meaningfully compare theoretical line profiles to observed TTS ones. The reasons are threefold: (a) velocity fields in TTS envelopes are not spherically symmetric and so require a 3D analysis; (b)  the atomic line level populations, which are dominated by non-LTE effects, must be computed iteratively with the radiation field; and (c) stellar envelopes are not isothermal. Point (a) can be addressed relatively easily by extending the present formalism to axisymmetric flow geometries. Point (b) requires setting up an iterative procedure that couples levels equations and the radiation field; several convergence acceleration methods have been proposed \citep[cf.][]{2009nrt..book..101A} and can be used to solve this problem. Point (c) is easily remedied; actually, a version of our code already makes provisions for varying envelope temperatures. Another aspect requiring improvement for further progress is the approximation used for the collisional de-excitation rates, which should be replaced by more recent data. This will be a direct consequence of setting up realistic atom models for addressing point (b).

\begin{acknowledgements}
It is a pleasure to thank Sylvie Cabrit for useful discussions and Joli Adams for improving the English. I am grateful to an anonymous referee for useful comments and suggestions that improved the paper's presentation.
\end{acknowledgements}

\bibliographystyle{aa.bst} % style aa.bst
\bibliography{49817corr.bib} % your references Yourfile.bib

\begin{thebibliography}{42}
\expandafter\ifx\csname natexlab\endcsname\relax\def\natexlab#1{#1}\fi

\bibitem[{{Armeni} {et~al.}(2023){Armeni}, {Stelzer}, {Claes}, {Manara},
  {Frasca}, {Alcal{\'a}}, {Walter}, {K{\'o}sp{\'a}l}, {Campbell-White},
  {Gangi}, {Mauco}, \& {Tychoniec}}]{2023A&A...679A..14A}
{Armeni}, A., {Stelzer}, B., {Claes}, R.~A.~B., {et~al.} 2023, \aap, 679, A14

\bibitem[{{Auer}(2009)}]{2009nrt..book..101A}
{Auer}, L.~H. 2009, in Numerical Radiative Transfer (Cambridge, UK: Cambridge
  University Press), 101

\bibitem[{{Bastian} {et~al.}(1980){Bastian}, {Bertout}, {Stenholm}, \&
  {Wehrse}}]{1980A&A....86..105B}
{Bastian}, U., {Bertout}, C., {Stenholm}, L., \& {Wehrse}, R. 1980, \aap, 86,
  105

\bibitem[{{Bertout}(1984)}]{1984ApJ...285..269B}
{Bertout}, C. 1984, \apj, 285, 269

\bibitem[{{Bertout} {et~al.}(1982){Bertout}, {Wolf}, {Carrasco}, \&
  {Mundt}}]{1982A&AS...47..419B}
{Bertout}, C., {Wolf}, B., {Carrasco}, L., \& {Mundt}, R. 1982, \aaps, 47, 419

\bibitem[{{Bouvier} {et~al.}(2007){Bouvier}, {Alencar}, {Harries},
  {Johns-Krull}, \& {Romanova}}]{2007prpl.conf..479B}
{Bouvier}, J., {Alencar}, S.~H.~P., {Harries}, T.~J., {Johns-Krull}, C.~M., \&
  {Romanova}, M.~M. 2007, in Protostars and Planets V, ed. B.~{Reipurth},
  D.~{Jewitt}, \& K.~{Keil}, 479

\bibitem[{{Bowen}(1934)}]{1934PASP...46..146B}
{Bowen}, I.~S. 1934, \pasp, 46, 146

\bibitem[{{Bowen}(1935)}]{1935ApJ....81....1B}
{Bowen}, I.~S. 1935, \apj, 81, 1

\bibitem[{{Cabrit} \& {Bertout}(1986)}]{1986ApJ...307..313C}
{Cabrit}, S. \& {Bertout}, C. 1986, \apj, 307, 313

\bibitem[{{Cabrit} \& {Bertout}(1990)}]{1990ApJ...348..530C}
{Cabrit}, S. \& {Bertout}, C. 1990, \apj, 348, 530

\bibitem[{{Cabrit} \& {Bertout}(1992)}]{1992A&A...261..274C}
{Cabrit}, S. \& {Bertout}, C. 1992, \aap, 261, 274

\bibitem[{{Castor}(1970)}]{1970MNRAS.149..111C}
{Castor}, J.~I. 1970, \mnras, 149, 111

\bibitem[{{Cox}(2000)}]{2000asqu.book.....C}
{Cox}, A.~N. 2000, {Allen's astrophysical quantities} (New York: AIP Press)

\bibitem[{{Dmitriev} {et~al.}(2019){Dmitriev}, {Grinin}, \&
  {Katysheva}}]{2019AstL...45..371D}
{Dmitriev}, D.~V., {Grinin}, V.~P., \& {Katysheva}, N.~A. 2019, Astronomy
  Letters, 45, 371

\bibitem[{{Eriksson} {et~al.}(2004){Eriksson}, {Veenhuizen}, {Wahlgren}, \&
  {Johansson}}]{2004RMxAC..21..132E}
{Eriksson}, M., {Veenhuizen}, H., {Wahlgren}, G.~M., \& {Johansson}, S. 2004,
  in Revista Mexicana de Astronomia y Astrofisica Conference Series, Vol.~21,
  Revista Mexicana de Astronomia y Astrofisica Conference Series, ed.
  C.~{Allen} \& C.~{Scarfe}, 132--136

\bibitem[{{Goldreich} \& {Kwan}(1974)}]{1974ApJ...189..441G}
{Goldreich}, P. \& {Kwan}, J. 1974, \apj, 189, 441

\bibitem[{{Gullbring} {et~al.}(1998){Gullbring}, {Hartmann}, {Brice{\~n}o}, \&
  {Calvet}}]{1998ApJ...492..323G}
{Gullbring}, E., {Hartmann}, L., {Brice{\~n}o}, C., \& {Calvet}, N. 1998, \apj,
  492, 323

\bibitem[{{Hamann}(1981)}]{1981A&A....93..353H}
{Hamann}, W.~R. 1981, \aap, 93, 353

\bibitem[{{Hartigan} {et~al.}(1995){Hartigan}, {Edwards}, \&
  {Ghandour}}]{1995ApJ...452..736H}
{Hartigan}, P., {Edwards}, S., \& {Ghandour}, L. 1995, \apj, 452, 736

\bibitem[{{Hartmann} {et~al.}(1982){Hartmann}, {Avrett}, \&
  {Edwards}}]{1982ApJ...261..279H}
{Hartmann}, L., {Avrett}, E., \& {Edwards}, S. 1982, \apj, 261, 279

\bibitem[{{Herbig}(1945)}]{1945PASP...57..166H}
{Herbig}, G.~H. 1945, \pasp, 57, 166

\bibitem[{Higgins(1898)}]{1898ApJ.....8R..54H}
Higgins, M. 1898, ApJ, 8

\bibitem[{{Hubeny} \& {Mihalas}(2014)}]{2014tsa..book.....H}
{Hubeny}, I. \& {Mihalas}, D. 2014, {Theory of Stellar Atmospheres} (Princeton:
  Princeton University Press)

\bibitem[{{Hyung} {et~al.}(2018){Hyung}, {Lee}, \& {Lee}}]{2018JASS...35....7H}
{Hyung}, S., {Lee}, S.-J., \& {Lee}, K.~H. 2018, Journal of Astronomy and Space
  Sciences, 35, 7

\bibitem[{{Joy}(1945)}]{1945ApJ...102..168J}
{Joy}, A.~H. 1945, \apj, 102, 168

\bibitem[{{Kastner}(1991)}]{1991Ap&SS.185..265K}
{Kastner}, S.~O. 1991, \apss, 185, 265

\bibitem[{{Kurosawa} \& {Romanova}(2012)}]{2012MNRAS.426.2901K}
{Kurosawa}, R. \& {Romanova}, M.~M. 2012, \mnras, 426, 2901

\bibitem[{{Lamers} {et~al.}(1987){Lamers}, {Cerruti-Sola}, \&
  {Perinotto}}]{1987ApJ...314..726L}
{Lamers}, H.~J.~G.~L.~M., {Cerruti-Sola}, M., \& {Perinotto}, M. 1987, \apj,
  314, 726

\bibitem[{{Olson}(1982)}]{1982ApJ...255..267O}
{Olson}, G.~L. 1982, \apj, 255, 267

\bibitem[{{Pavlakis} \& {Kylafis}(1996)}]{1996ApJ...467..292P}
{Pavlakis}, K.~G. \& {Kylafis}, N.~D. 1996, \apj, 467, 292

\bibitem[{{Pereira} {et~al.}(1999){Pereira}, {de Ara{\'u}jo}, \&
  {Landaberry}}]{1999MNRAS.309.1074P}
{Pereira}, C.~B., {de Ara{\'u}jo}, F.~X., \& {Landaberry}, S.~J.~C. 1999,
  \mnras, 309, 1074

\bibitem[{Rybicki \& Hummer(1978)}]{1978ApJ...219..654R}
Rybicki, G.~B. \& Hummer, D.~G. 1978, ApJ, 219, 654

\bibitem[{{Selvelli} {et~al.}(2007){Selvelli}, {Danziger}, \&
  {Bonifacio}}]{2007A&A...464..715S}
{Selvelli}, P., {Danziger}, J., \& {Bonifacio}, P. 2007, \aap, 464, 715

\bibitem[{{Sobolev}(1960)}]{1960mes..book.....S}
{Sobolev}, V.~V. 1960, {Moving envelopes of stars} (Cambridge: Harvard
  University Press)

\bibitem[{{Tambovtseva} {et~al.}(2001){Tambovtseva}, {Grinin}, {Rodgers}, \&
  {Kozlova}}]{2001ARep...45..442T}
{Tambovtseva}, L.~V., {Grinin}, V.~P., {Rodgers}, B., \& {Kozlova}, O.~V. 2001,
  Astronomy Reports, 45, 442

\bibitem[{{Tambovtseva} {et~al.}(2014){Tambovtseva}, {Grinin}, \&
  {Weigelt}}]{2014A&A...562A.104T}
{Tambovtseva}, L.~V., {Grinin}, V.~P., \& {Weigelt}, G. 2014, \aap, 562, A104

\bibitem[{{Thackeray}(1937)}]{1937ApJ....86..499T}
{Thackeray}, A.~D. 1937, \apj, 86, 499

\bibitem[{{Valenti} {et~al.}(1993){Valenti}, {Basri}, \&
  {Johns}}]{1993AJ....106.2024V}
{Valenti}, J.~A., {Basri}, G., \& {Johns}, C.~M. 1993, \aj, 106, 2024

\bibitem[{{van Regemorter}(1962)}]{1962ApJ...136..906V}
{van Regemorter}, H. 1962, \apj, 136, 906

\bibitem[{{Willson}(1972)}]{1972A&A....17..354W}
{Willson}, L.~A. 1972, \aap, 17, 354

\bibitem[{{Willson}(1974)}]{1974ApJ...191..143W}
{Willson}, L.~A. 1974, \apj, 191, 143

\bibitem[{{Willson}(1975)}]{1975ApJ...197..365W}
{Willson}, L.~A. 1975, \apj, 197, 365

\end{thebibliography}

\appendix
\section{Numerical implementation \label{sec:SLIM2}}

\textit {SLIM2} (Spectral Line Interactions in Moving Media) is a publicly available\footnote{under Creative Commons 4.0 BY-NC-SA license. The code is available at \url{https://github.com/claude-bertout/SLIM2}}  Python/NumPy code for solving the line formation problem in 2D moving media, for multiple lines stemming from multiple elements, in the special case of interacting lines leading to fluorescence effects. In the framework of the well-tested hybrid approach to the problem of line formation described above, it first computes the generalized Sobolev source functions and then integrates the line flux exactly .

The code makes extensive use of the NumPy ndarray((dim1, dim2), object) 3D construct, which is particularly useful for ray-tracing procedures, although it is frowned upon by some pythonistas because if complicates array broadcasting. 

Version 1.0 presented here is stable and optimized for solving the problem of forming the three-line blend around 3969\AA\ in TTS spectra. The atomic level populations are LTE and remain constant during the iterative procedure used to compute the source functions. Plotting routines are restricted to this specific three-line case. They are included in the project so that interested users will be able to reproduce easily this publication's figures, thus gaining experience with the code before starting on their own developments. 

A more general version of the code, which is able to compute many interacting and noninteracting lines simultaneously and allows for variable temperature in the envelope, will be made available at a later date.

\smallskip
\noindent The following modules are included in the GitHub TTS Fluorescence package:
\begin{itemize}
        \item [\textbullet] \textbf {Fluorescence\shortunderscore  Main.py} contains a single function \textbf {main\shortunderscore  module} that calls all other modules in the following order to perform the computation and to store computational parameters, source functions, and line profiles in a subdirectory of \textbf {results} directory marked with a date-time stamp of the start of computation.
        \item  [\textbullet] \textbf {import\shortunderscore  parameters.py} imports all input parameters from \textbf {computation\shortunderscore  parameters.conf}. For this, an ancillary module \textbf {parameter\shortunderscore  parser.py} is needed. Several different options are possible for the computation (local or nonlocal Sobolev source, noninteracting or interacting lines, Sobolev or exact flux mode, single model or grid of models, various velocity fields, several levels of logging,  etc.) so that functional parameters are numerous. There are also a few physical parameters, such as the maximum density, envelope temperature, and stellar properties, that must be defined. The meaning of each of these parameters is explained in the .conf file and default values are indicated.
        \item  [\textbullet] \textbf {setup\shortunderscore  grids.py} sets up the spatial, velocity, and density grids and analyzes the velocity field structure.
        \item [\textbullet] \textbf {import\shortunderscore  line\shortunderscore  data.py} reads the line data from \textbf {import\shortunderscore  atomic\shortunderscore  data.py} and computes and stores as arrays the line optical depths using \textbf {LTE\shortunderscore  module.py}. It also computes the absorption line profiles, defines the line interaction matrices, and sets up the overall frequency grid for the flux integration. Databases for partition functions and abundances are stored in directory \textbf {input\shortunderscore  data}.
        \item  [\textbullet] \textbf {source\shortunderscore  module.py} first finds and orders all the CP surfaces discussed above on high-resolution angular grids. This step is a major CPU-time consumer, and we store the CP-surface geometries for repeated use when a string of models with the same geometrical properties are calculated. All integrals over solid angles that appear in the computation of the source function are solved by Gaussian quadratures. The local source functions are computed and iterated with the nonlocal contributions until convergence occurs. Although there are often more than two resonant surfaces on each line of sight, convergence occurs relatively rapidly for the velocity fields considered above and for the many others that we tested. 
        % The \citet{1978ApJ...219..654R} theorem about convergence in the case of two interacting surfaces can probably be generalized to multiple surfaces, 
        % at least in the LTE case that we considered here.
        \item [\textbullet] \textbf {flux\shortunderscore  integration.py} integrates the line intensities over the impact parameter grid. A spatial grid is defined on each impact parameter and is transformed by interpolation into a grid of line frequencies present on the given ray. Contributing frequencies to each line are then found by taking their velocity displacements into account, and resonant regions are deduced from there. Incremental line intensities are then computed, and the flux follows from integrating the emergent intensities.
        \item  [\textbullet] \textbf {function\shortunderscore  library.py} contains ancillary functions for \textbf {source\shortunderscore  module.py} and \textbf {flux\shortunderscore  integration.py}.
        \item  [\textbullet] \textbf{in\shortunderscore  out\shortunderscore  library.py} contains functions for storing the results in a subdirectory of \textbf{results} and reading them for future use.
        \item  [\textbullet] Finally, \textbf {plot\shortunderscore  library.py} reads stored results and plots various quantities, storing the plots for future reference in the same subdirectory of \textbf {results} (see above).
\end{itemize}   

All these modules are heavily commented to make it easier for the user to follow the procedures used for solving the problem. 

The computing time for three interacting lines with high-resolution grids (100 radial points, 2000 angles, 100 impact parameters, and 2000 frequencies across the line profiles) is typically on the order of 30 seconds on an Apple MacBook Pro with M1 Pro silicon running PyCharm. 

\end{document}